\documentclass[conference]{IEEEtran}
\IEEEoverridecommandlockouts
\usepackage{cite}
\usepackage{amsmath,amssymb,amsfonts}
\usepackage{algorithmic}
\usepackage{graphicx}
\usepackage{textcomp}
\usepackage{xcolor}
\usepackage[hyphens]{url}
\usepackage{fancyhdr}
\usepackage[bookmarks=true,breaklinks=true,letterpaper=true,colorlinks,citecolor=blue,linkcolor=blue,urlcolor=blue]{hyperref}

\def\BibTeX{{\rm B\kern-.05em{\sc i\kern-.025em b}\kern-.08em
    T\kern-.1667em\lower.7ex\hbox{E}\kern-.125emX}}
    \makeatletter
\newcommand{\linebreakand}{%
  \end{@IEEEauthorhalign}
  \hfill\mbox{}\par
  \mbox{}\hfill\begin{@IEEEauthorhalign}
}

\makeatother
\begin{document}

\title{HgPCN: A Heterogeneous Architecture for E2E Embedded Point Cloud Inference
%{\footnotesize \textsuperscript{*}Note: Sub-titles are not captured for https://ieeexplore.ieee.org  and should not be used}
\thanks{This research was supported in part by the National Science Foundation (NSF) Center for Space, High-Performance, and Resilient Computing (SHREC) through the IUCRC Program under Grant No. CNS-1738420.}
}

\author{\IEEEauthorblockN{1\textsuperscript{st} Yiming Gao}
\IEEEauthorblockA{%\textit{Department of Electrical \& Computer Engineering} \\
\textit{University of Florida}\\
gaoyiming@ufl.edu}
\and
\IEEEauthorblockN{2\textsuperscript{nd} Chao Jiang}
\IEEEauthorblockA{%\textit{{Department of Electrical \& Computer Engineering}} \\
\textit{University of Florida}\\
jc19chaoj@ufl.edu}
\and
\IEEEauthorblockN{3\textsuperscript{rd} Wesley Piard}
\IEEEauthorblockA{%\textit{dept. name of organization (of Aff.)} \\
\textit{University of Florida}\\
wespiard@ufl.edu}

\linebreakand

\IEEEauthorblockN{4\textsuperscript{th} Xiangru Chen}
\IEEEauthorblockA{%\textit{dept. name of organization (of Aff.)} \\
\textit{University of Florida}\\
cxr1994816@ufl.edu}
\and
\IEEEauthorblockN{5\textsuperscript{th} Bhavesh Patel}
\IEEEauthorblockA{%\textit{dept. name of organization (of Aff.)} \\
\textit{Dell EMC}\\
Bhavesh.A.Patel@dell.com}
\and
\IEEEauthorblockN{6\textsuperscript{th} Herman Lam\textsuperscript{*}}
\IEEEauthorblockA{%\textit{{Department of Electrical \& Computer Engineering}} \\
\textit{University of Florida}\\
hlam@ufl.edu}
}

\maketitle

\begin{abstract}

Point cloud is an important type of geometric data structure for many embedded applications such as autonomous driving and augmented reality. Current Point Cloud Networks (PCNs) have proven to achieve great success in using inference to perform point cloud analysis, including object part segmentation, shape classification, and so on. However, point cloud applications on the computing edge require more than just the inference step. They require an end-to-end (E2E) processing of the point cloud workloads: pre-processing of raw data, input preparation, and inference to perform point cloud analysis.  Current PCN approaches to support end-to-end processing of point cloud workload cannot meet the real-time latency requirement on the edge, i.e., the ability of the AI service to keep up with the speed of raw data generation by 3D sensors.

Latency for end-to-end processing of the point cloud workloads stems from two reasons: memory-intensive down-sampling in the pre-processing phase and the data structuring step for input preparation in the inference phase. In this paper, we present HgPCN, an end-to-end heterogeneous architecture for real-time embedded point cloud applications. In HgPCN, we introduce two novel methodologies based on \emph{spatial indexing} to address the two identified bottlenecks. In the Pre-processing Engine of HgPCN, an Octree-Indexed-Sampling method is used to optimize the memory-intensive down-sampling bottleneck of the pre-processing phase. In the Inference Engine, HgPCN extends a commercial DLA with a customized Data Structuring Unit which is based on a Voxel-Expanded Gathering method to fundamentally reduce the workload of the data structuring step in the inference phase.

The initial prototype of HgPCN has been implemented on an Intel PAC (Xeon+FPGA) platform. Four commonly available point cloud datasets were used for comparison, running on three baseline devices: Intel Xeon W-2255, Nvidia Xavier NX Jetson GPU, and Nvidia 4060ti GPU. These point cloud datasets were also run on two existing PCN accelerators for comparison: PointACC and Mesorasi. Our results show that for the inference phase, depending on the dataset size, HgPCN achieves speedup from 1.3$\times$ to 10.2$\times$ vs. PointACC, 2.2$\times$ to 16.5$\times$ vs. Mesorasi, and 6.4$\times$ to 21$\times$ vs. Jetson NX GPU. Along with optimization of the memory-intensive down-sampling bottleneck in pre-processing phase, the overall latency shows that HgPCN can reach the real-time requirement by providing end-to-end service with keeping up with the raw data generation rate.

\end{abstract}


\section{Introduction}
\label{Introduction}
With the development of 3D sensors, such as LiDARs, and RGB-D cameras, there are increased interests in research and development of processing point cloud data. Point Cloud Networks (PCNs) have proven to achieve great success in different tasks of point cloud analysis, including object part segmentation, shape classification, and so on.
Point cloud applications on the computing edge, such as autonomous robotics and drones, have stringent real-time requirements. Many PCNs for point cloud AI tasks have been proposed \cite{guo2020deep}, including DSA (Domain-Specific Architecture) PCN accelerators such as Mesorasi \cite{9251968}, PointACC \cite{lin2021pointacc}, and Crescent \cite{feng2022crescent}, which have been proposed to accelerate PCN inference. However, point cloud applications on the computing edge require more than just the inference step. They require an end-to-end processing of the point cloud workloads. A recent paper \cite{9065599} indicated that the processing of AI services for edge computing suffer the problem of high “AI tax”, which includes the supporting steps for AI workload such as pre-processing of raw data, input preparation, and offloading communication overhead. These steps contribute to a major part of overall latency in edge AI services. Unfortunately, this high “AI tax” problem is more serious for point cloud applications. Due to the inherent irregularity and large size of raw point cloud data, the pre-processing and communication overhead of point cloud data are far more expensive than traditional data types, such as image or video. Due to the high AI-tax, the current PCN approaches to support end-to-end processing of point cloud workload cannot meet the real-time requirement \cite{hu2020randla}, i.e., the ability of the AI service to keep up with the speed of raw data generation by 3D sensors. In PCN edge services, the AI tax is mainly stemmed from two reasons: the expensive down-sampling pre-processing phase and the data structuring step for input preparation in the inference phase.

\begin{figure}[t]
% \centerline{\includegraphics{intro_v2_output.png}}
\centerline{\includegraphics{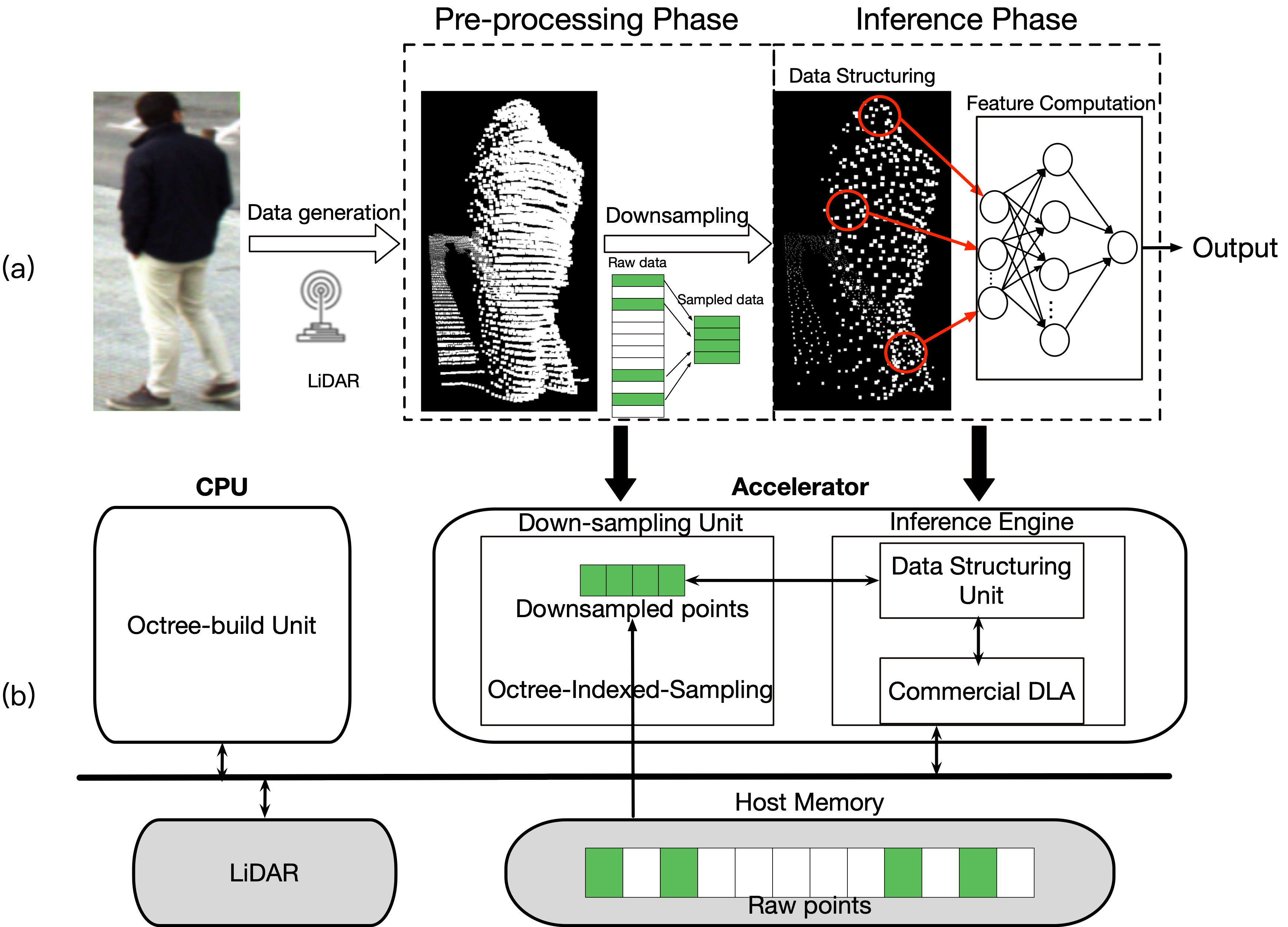}}

\caption{(a) Two phases of an end-to-end point clouds AI service (classification task), (b) Overall architecture to process the two phases.}
  \label{fig.intro}

\end{figure}

As shown in Figure~\ref{fig.intro}(a), an end-to-end point cloud AI service based on PCNs comprise of two major phases after raw data generation: pre-processing of the point cloud data and PCN inference. After obtaining the raw point cloud data from the sensor, the size of the raw data is enormous: e.g., for every frame, a LiDAR sensor produces approximately 2 million points. Also, the number of points in each frame is highly irregular because different objects have different reflectivity to the laser. Thus, point cloud applications on the edge require an effective front-end pre-processing phase to deal with this kind of irregularity before feeding the data into the PCN for inference. For each frame, it has to be down-sampled from million-level, variable-number points into thousands-level, fixed number points (e.g., 4096 points per frame to feed into the input layer of a PCN). Because of the large amounts of points in raw point cloud data and expensive nature of down-sampling methods, down-sampling pre-processing is extremely memory intensive and is \noindent\textbf{a major bottleneck in a real-time point cloud application.}

As a type of Deep Neural Network (DNN), the backend PCN inference phase is a computationally intensive task. In recent years, there have been much advance in Deep Learning Accelerators (DLAs), including commercially available DLAs \cite{chen2020survey}, which perform well in satisfying the computationally intensive inference requirements of traditional DNNs. However, current DLAs cannot be directly applied to PCN inference. The reason is, unlike traditional dense data types, points in a point cloud are sparsely distributed in a 3D space \cite{xu2020grid, guo2020deep}. Thus, before the actual feature computation with MVM (matrix-vector multiplication), an additional step, \textbf{data structuring}, is necessary to adapt the spatial sparsity of the point cloud and prepare the input for the following convolutional layers. The data structuring step is not supported by current DLAs. However, if unaccelerated, this step results in a non-trivial part of the total computation of PCN inference. Data structuring before feature computation is the \textbf{second major bottleneck in a real-time point cloud application}.

Current general-purpose architectures fail to effectively solve these two bottlenecks of end-to-end point cloud AI services for real-time applications. For example, the most commonly used down-sampling pre-processing method, farthest-point sampling (FPS), takes over 200 seconds to sample 10\% of 1 million points \cite{hu2020randla} on GPUs. For CPUs, such enormous amount of computation is rather slow because their limited parallel capability. We will show in Section \ref{Motivation} that the down-sampling pre-processing phase occupies a major part of the end-to-end latency, which greatly exacerbating the problem of high “AI tax” \cite{9065599}.

Currently, there is a very limited number of hardware-based designs for point cloud accelerators \cite{9251968, lin2021pointacc}. Furthermore, these works do not provide a complete solution since they do not fully consider the two bottlenecks. They focus solely on the AI (inference) phase, based on the assumption that the input is already down-sampled (pre-processed) and restructured. More details on these previous works will be given in Section \ref{Background}.  

In this paper, we first identify the workload characteristics and bottlenecks of the two major phases of end-to-end point cloud edge applications. The frontend pre-processing phase is a memory-intensive task, whereas the backend inference phase is a computationally intensive task. The workload characteristics are discussed in Section \ref{Motivation}. An overall architecture for HgPCN to support end-to-end processing of point-cloud applications is presented in Section \ref{Architecture}.

To enable the real-time PCN process in HgPCN, we introduce two novel methodologies to replace the brute-force traditional pre-processing and data structuring methods (introduced in Section \ref{Background}) by converting them into efficient Octree-based spatial query.
To address the memory-intensive bottleneck in the pre-processing phase, we develop an \emph{Octree-Indexed-Sampling} (OIS) method. The OIS method uses an Octree data structure as a \textit{spatial} index \cite{schon2013octree} to organize the point cloud data in memory. With an Octree pre-organization, down-sampling can be done with only reading out the desired after-sampled points among raw point cloud data from the memory, eliminating the need for repetitive access of input and intermediate data from memory. In the HgPCN architecture, the OIS method is implemented in the \textbf{Pre-processing Engine} shown in Figure~\ref{fig.intro}(b), which comprised of the Octree-Build Unit in the CPU and the Down-sampling Unit in the accelerator (e.g., FPGA). The design and implementation of the Pre-processing Engine are described in detail in Section \ref{Pre-processing engine}.

For the Inference phase, we develop an Octree-based \emph{Voxel-Expanded Gathering} (VEG) method to optimize the data structuring step.
In our prototype implementation, the \textbf{Inference Engine} (shown in Figure~\ref{fig.intro}(b)) will make use of a commercially available DLA. It will be enhanced by a Data Structuring Unit (DSU) to form the PCN Inference Accelerator. As described in more detail in Section \ref{Inference engine}, the Data Structuring Unit is based on the VEG method and fundamentally reduces the workload of the data structuring step as compared against current methods.
  \begin{figure*}[t]
\centering
\includegraphics[width=1\linewidth]{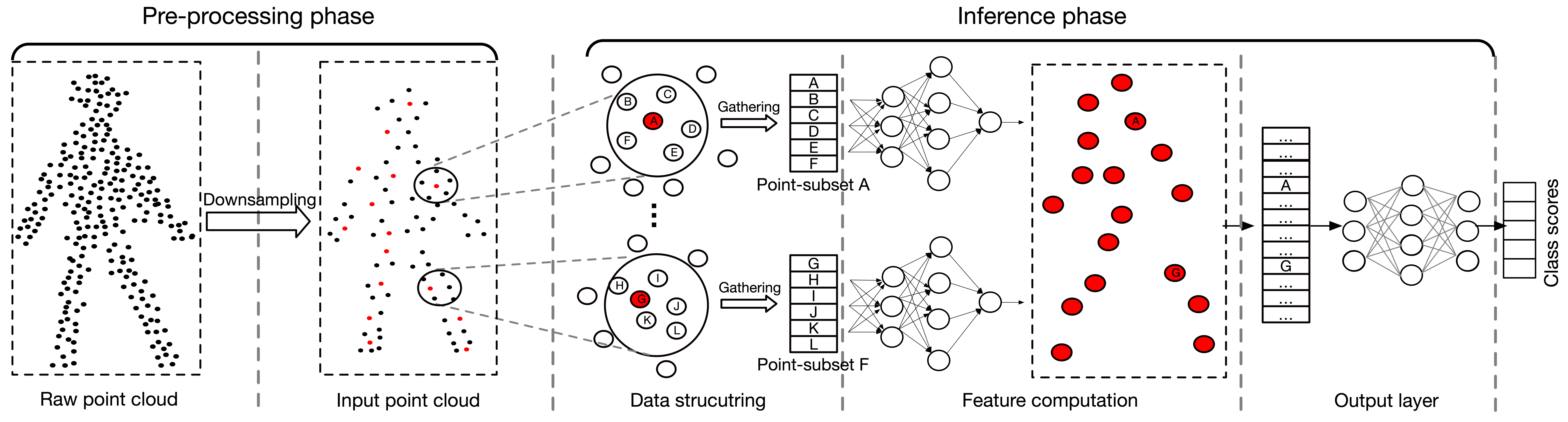}
\caption{Illustration of the steps of an end-to-end PCN inference (toy-valued pedestrian classification task).}
\label{diffratioinhybrid}
\label{fig.background_pre_final}
\end{figure*} \normalsize

The initial prototype of HgPCN has been implemented on an Intel PAC (Xeon+FPGA) platform. The evaluation of the HgPCN prototype is performed using four modern point cloud datasets: ModelNet40 \cite{sunmodelnet40}, ShapeNet \cite{chang2015shapenet}, S3DIS \cite{landrieu2018large }, and KITTI \cite{geiger2013vision}. The comparing baseline includes CPU/GPUs, and existing DSA PCN accelerators: PointACC\cite{lin2021pointacc}, Mesorasi\cite{9251968}. The evaluation of HgPCN is first performed independently for the two major phases (from engine-level), then evaluate the E2E latency from the HgPCN system-level in the edge computing scenario. The description of the prototype HgPCN and its evaluation are detailed in Section \ref{Evaluation}.

In Section~\ref{Conclusions}, we provide the conclusions of our work and a discussion of our future directions.

\section{Background and Related Work}
\label{Background}

\subsection{Point Cloud Data and PCNs}

 A point cloud is a set of points $x = \left\{(p_{k}, f_{k})\right\}$, where $p_k = (x_{k}, y_{k}, z_{k})$ is the coordinate of the $k^{th}$ point, and $f_{k}$ is the corresponding 1-D feature vector. Unlike a traditional image, which is a dense 2D pixel-matrix, a point cloud is comprised of numerous spatially distributed points in a 3D space. Point clouds are commonly generated by 3D data acquisition devices (such as LiDAR) and generally are massive and ever-changing. For example, in the KITTI dataset \cite{geiger2013vision}, each frame includes approximately $(N {\sim} 10^6)$ to $(N {\sim} 10^7)$ points; and the number of points varies widely between frames. Current DLAs processing DNNs cannot consume such large and irregular input datasets. Thus, a pre-processing (down-sampling) step is necessary. As shown in Figure~\ref{fig.background_pre_final}, the down-sampling step is used to decimate the original point cloud into a fixed number of points, while maintaining the spatial information structure.

The down-sampled input point cloud is the input to the backend PCN inference phase, as shown in Figure~\ref{fig.background_pre_final}. Similar to the concept of a “stride window” in convolution, weight kernels in PCN will only be applied to a subset of input points while moving over the entire point set. Because point clouds are sparsely scattered in a 3D space, forming the input subset of points is far more difficult than traditional dense input like images. Unlike the pixels in images, there is no direct neighbor-indexing method in point clouds. Thus, as shown in Figure~\ref{fig.background_pre_final}, before the feature computation step, an extra data structuring step is required to form the subset of points as the “input feature map” by using neighbor-gathering methods.

The feature computation step for inference of PCNs in Figure~\ref{fig.background_pre_final} makes use of traditional DNNs outputted from the data structuring step. The weight will be applied to the input points subset in the MLP hidden layer and the final outputs are obtained from the output layer. The feature computation step can be decomposed into MVM, and can be directly accelerated by existing commercially avaibable DLAs such as NPU \cite{esmaeilzadeh2014neural}.

In summary, an end-to-end PCN inference process contains three major steps: down-sampling pre-processing of the raw point cloud data, data structuring, and feature computation. Among these operations, down-sampling and data structuring are point-cloud-specific operations and have not been fully optimized. However, these two operations are quite expensive and become major bottlenecks for end-to-end PCN inference.

\noindent\textbf{Expensive downsampling pre-processing.} In real-world edge computing applications, pre-processing is a necessary step in the AI pipeline \cite{9065599}. Furthermore, for point clouds, the pre-processing is far more expensive than traditional data types and contributes to a major part of the latency. There are two main causes of the high latency. The first is the enormous size of raw point cloud data. The number of points in raw point cloud data is from $N {\sim} 10^5$ to $N {\sim} 10^6$. The second reason is the expensive down-sampling methods used. For example, the FPS method samples the raw point cloud iteratively. It starts by randomly selecting a seed point from the raw point cloud set (denoted as set  $C$) and putting it in the sampled points set (denoted as set  $S$). In each iteration, it picks a point from the unpicked point set $C-S$ that is the farthest from the sampled point set $S$ and adds it into the sampled point set $S$, until $S$ contains the predefined number of points $K$. In this way, the FPS method will return a subset $\left\{ p^{1}, {\cdots}, p^{k}, {\cdots}, p^{K} \right\}$ from raw points $C$, such that each $p^{k}$ is the farthest point from the first $k - 1 $ points. $K$ is a selected fixed number (e.g., 4096). We will show later in Section \ref{Motivation} that the down-sampling pre-processing phase is a memory-intensive task and is a huge computational workload. In addition, the down-sampling pre-processing phase does not use memory bandwidth efficiently. In the FPS method, over 99\% of memory accesses are wasted because most of these points will be filtered out and not be used again after down-sampling. Among all the existing sampling methods, only the Random Sampling (RS) method is possibly fast enough to satisfy real-time requirements on general-purpose architectures. The RS method simply picks $K$ points from the raw point cloud randomly. As a result, the accuracy of random sampling is low and cannot be fully trusted \cite{hu2020randla}, especially in some safety-oriented applications.

\noindent\textbf{Expensive Data Structuring.}
 As mentioned, in order to use commercially available DLAs, PCNs need to gather point-subsets, which are in the neighboring set $N(x^{k})$ of selected central points (red points in the Figure~\ref{fig.background_pre_final}), to form the “input feature map” for feature computation. (For example, in Figure~\ref{fig.background_pre_final}, central point A gathers nearest points B, C, D, E, and F to form the point-subset A.) Unlike pixels in images, the points in point clouds are scattered over a 3D space in an irregular manner. There is no direct neighbor-indexing method to gathering the neighboring set $N(x^{k})$. The gathering process can be achieved by expensive nearest neighbor gathering, such as KNN (K-nearest-neighbors) and BQ (Ball query). In traditional methods, neighbor points gathering for each central point needs to search over the entire input point cloud. Let’s assume the size of the input point cloud is $n$, and neighbor gathering size is $k$. For neighbor gathering of each central point, it is necessary to compute the distances from this central point to every other $n-1$ points, and pick the top $k$ points with nearest distances. Each neighbor gathering process of a central point replicates these steps and results in a high workload.
 
\subsection{Current PCN Accelerators}

Existing PCN accelerators mainly focus on optimizing the expensive data structuring step (second bottleneck), with the assumption the point cloud data is already pre-processed (first bottleneck). Unlike these PCN accelerators, the proposed HgPCN will not assume the point cloud data has been pre-processed. The HgPCN architecture supports efficient online pre-processing, together with an Inference Engine to supports efficient end-to-end PCN inference service.

To perform the data structuring step in the inference phase, existing PCN accelerators can be divided into two types. The first type \cite{lin2021pointacc, 9251968} performs data structuring by using  \textit{accurate neighbor search} methods, which returns the same result as traditional data structuring methods. The second type is to improve the latency of the data structuring step by using some tree-based methods \cite{feng2022crescent, xu2019tigris, pinkham2020quicknn}, to perform an  \textit{approximate neighbor search}. As a result, these approximation methods require some adaptation in the model training phase \cite{feng2022crescent}. The data structuring method used in HgPCN results in accurate (not approximate) data structuring, which is compatible with current training method. Thus, we will compare our inference results (in Section \ref{Evaluation}) to the first type of PCN accelerators.

\section{Motivation}
\label{Motivation}

In this section, we identify the key performance bottlenecks of an end-to-end PCN in general (not considering accelerators). The evaluation was performed using general-purpose platforms: Intel® Xeon® W-2255 CPU and 4060Ti GPU. For down-sampling in the pre-processing phase, the most commonly used FPS \cite{eldar1997farthest} method was used. For the backend, Pointnet++ \cite{qi2017pointnet++} was used to perform inference on the Modelnet40 \cite{sunmodelnet40}, ShapeNet \cite{chang2015shapenet}, S3DIS \cite{landrieu2018large }, and KITTI \cite{geiger2013vision} datasets. The end-to-end PCN application was run to obtain a breakdown of the latency. Then, we quantitatively analyze the detailed operations in the two phases to identify the bottlenecks which motivate the HgPCN design.
	\begin{figure}[h]
		\centerline{\includegraphics[width=0.99\linewidth]{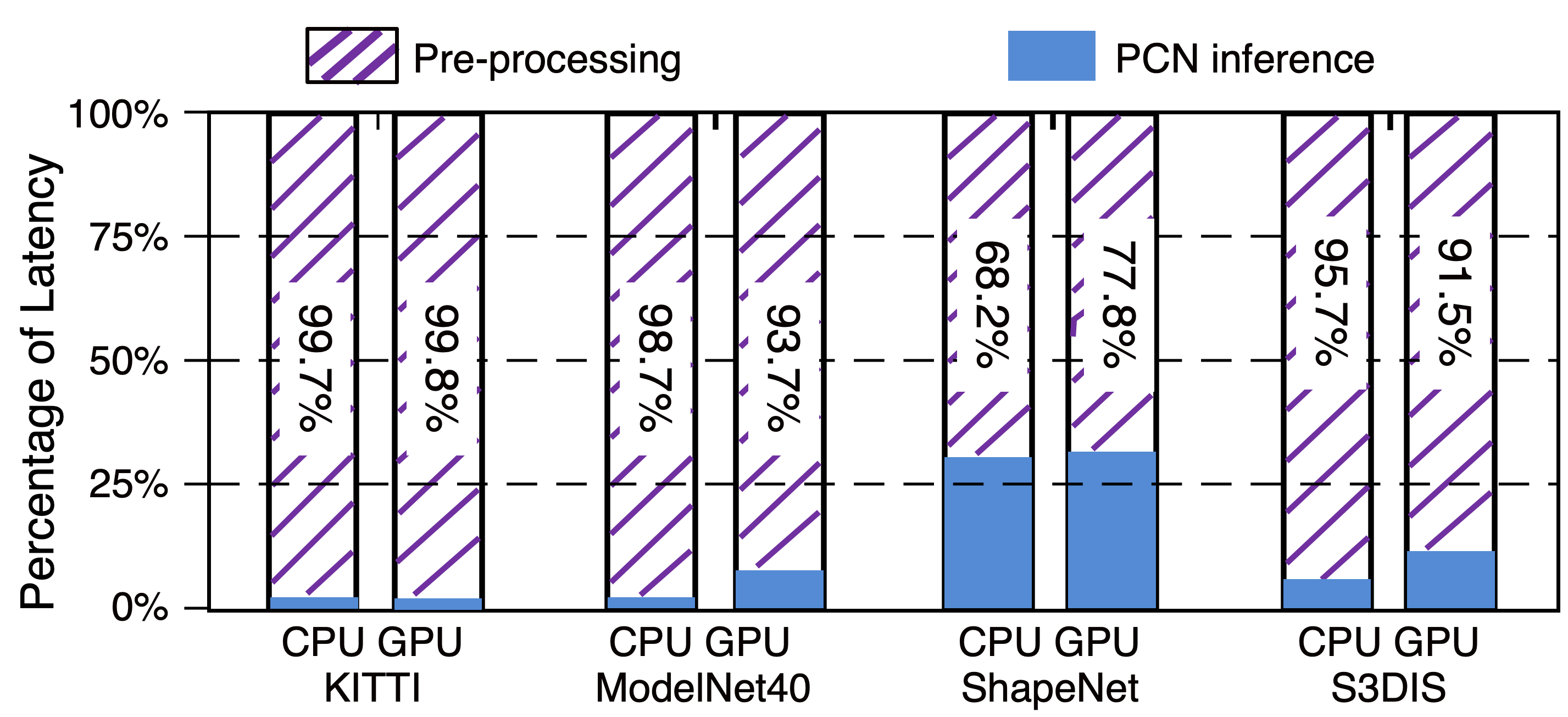}}
		\caption{End-to-end execution time breakdown (actual time not shown).}
	    \label{fig.overall_breakdown}
	\end{figure}
 
\noindent\textbf{Overall Latency Analysis.}
Figure~\ref{fig.overall_breakdown} shows the percentage of the total end-to-end latency spent on the pre-processing phase and inference phase for each of the four example datasets.  As expected, datasets with larger raw point clouds require more time for pre-processing. From Figure~\ref{fig.overall_breakdown}, it can be observed that the latency of pre-processing is far greater than the latency of the actual inference. The results confirm that the high cost of the AI tax for point cloud applications is far more serious than traditional AI applications \cite{9065599}. Optimizing only the PCN inference phase, as done in existing PCN Accelerators \cite{jouppi2017datacenter}, is far from enough to meet the real-time requirements of point cloud applications on the computing edge.

\subsection{Analysis of Frontend Pre-processing}% (downsampling)}

\noindent\textbf{Bottleneck Identification.} 
The bottleneck in the pre-processing phase is caused by large data movement, resulting from the large memory footprint of the down-sampling process. One reason is the large input size. In our baseline datasets, for every frame of raw point cloud data, KITTI contains $(N {\sim} 10^{6})$, Modelnet40 contains $(N {\sim} 10^{5})$, and S3DIS contains $(N {\sim} 10^{5})$ points, respectively. The second reason is that an excessive amount of intermediate data is generated during the down-sampling phase. As we discussed in Section \ref{Background}, the FPS algorithm samples the raw point cloud by iteratively picking the farthest point. For each iteration, we have to compute the distances between every point in unpicked points set $C-S$ to picked points set $S$ and rank these distances. This process exhibits low data locality because all of the computed distances (intermediate data) are written into the memory, and then read again after all distances are calculated. For these reasons, the down-sampling phase is a very time-consuming memory-intensive task, and the major resource of AI tax. 

\subsection{Analysis of Backend PCN Inference}

As shown in Figure~\ref{fig.background_pre_final}, PCN inference contains two separate steps: data structuring (DS) and feature computation (FC). The DS step is used as input preparation for the FC step, which is the actual convolution (CONV) step.

\noindent\textbf{Bottleneck Identification.} 
The data structuring step consists of the neighbor search of central points, which can be solved by algorithms such as k-nearest-neighbors (KNN) or Ball query (BQ). Even though the neighbor search operations themselves do not involve complicated computations, it is still time-consuming because the neighbor search algorithms operate over the entire input point cloud. For the data structuring step of a point-subset of PCNs, we need to compute the distance from every point to the central point and pick the top k nearest points by ranking these distances (using the KNN example). And this data structuring step is replicated for every central point in the input point cloud. As a result, a large amount of computation is required by the data structuring steps for each feature computation step. A recent work \cite{lin2021pointacc} shows that in PointNet++, approximately 50\% of total computation of the inference phase is consumed by the data structuring step and becomes another source of AI tax.

The feature computation in PCNs (i.e., actual inference) shares the same requirements as the traditional DNNs and can be directly optimized by using DLAs \cite{jouppi2017datacenter}. But the data structuring step is unique for PCNs and are not supported by existing commercial accelerators. To take advantage of commercial DLAs to accelerating the feature computation step, the data structuring bottleneck needs to be addressed.

\section{HgPCN Architecture}
\label{Architecture}

\begin{figure}[t]
\centerline{\includegraphics[width=0.99\linewidth]{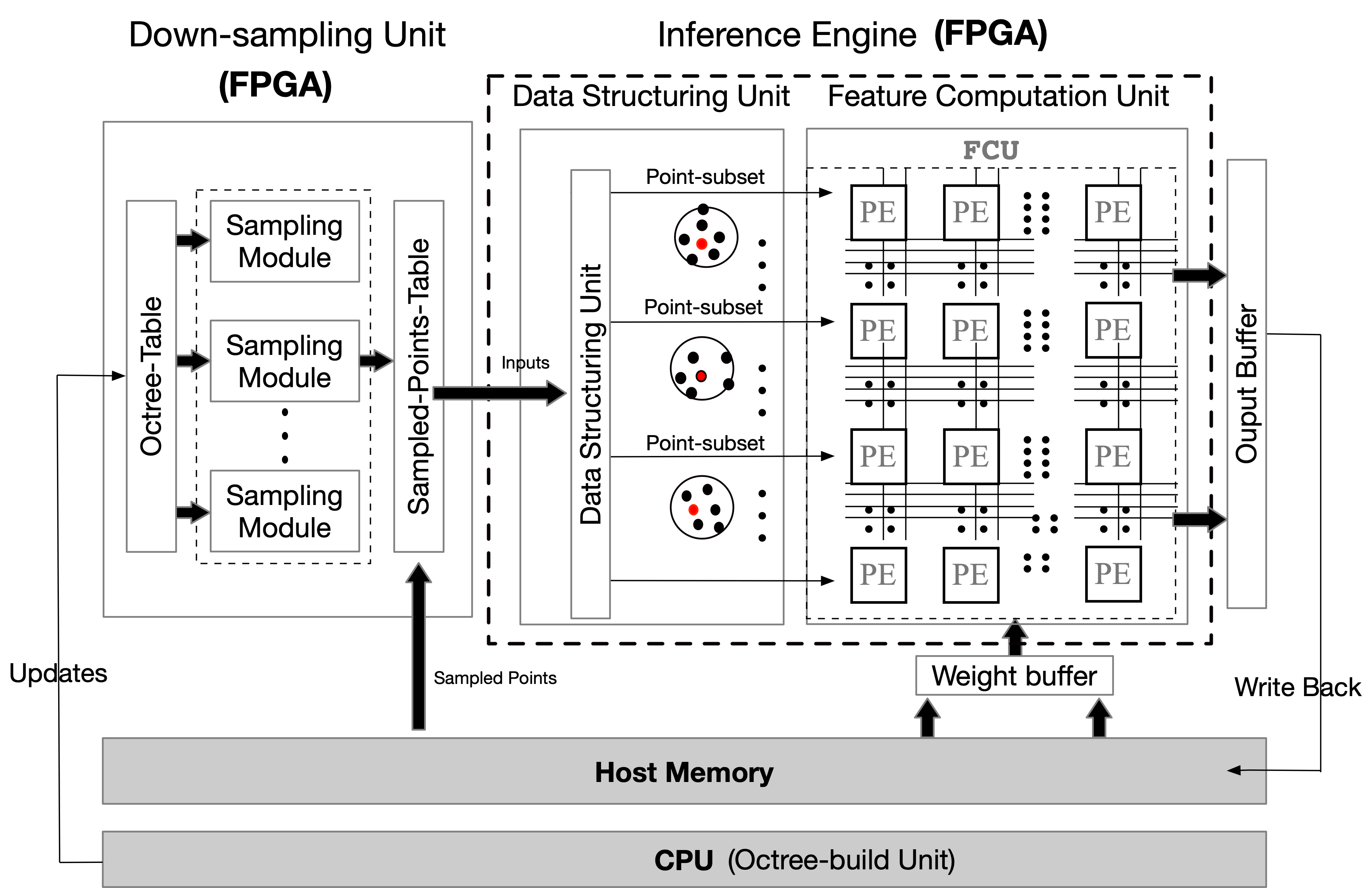}}
		\caption{Architecture overview of HgPCN.}
		\label{fig.architecture_paper_subset}
		\vspace{-3mm}
	\end{figure}

Shown in Figure~\ref{fig.architecture_paper_subset} is an overview of the architecture of HgPCN, which is based on a CPU-FPGA shared memory platform (e.g., Intel PAC card \cite{jiang2019acceleration}). The major system components of the architecture are: (1) CPU, (2) Host Memory, and (3) FPGA. The major architectural components of HgPCN  are (1) Pre-processing Engine (Octree-build Unit in the CPU and Down-sampling Unit in the FPGA) and (2) Inference Engine (Data Structuring Unit and Feature Computation Unit) in the FPGA. A key feature of the HgPCN architecture is that both the CPU and FPGA can access the data in the shared Host Memory. For each frame of the point cloud, we assume the raw data from the data sensor (e.g., LiDAR) is collected and stored in the Host Memory. Recall from Section \ref{Introduction} (Figure~\ref{fig.intro}), an end-to-end point cloud AI service based on PCNs comprise of two major phases after raw data generation: pre-processing of the point cloud data and PCN inference. As will be described in more detail later, for the HgPCN, the pre-processing phase consists of two steps; (1) an Octree construction \& memory pre-configuration step which is performed using the CPU and Host Memory, and (2) a down-sampling step which is performed in the Down-sampling Unit in the FPGA. The PCN inference phase is performed in the Inference Engine in the FPGA. As shown in Figure~\ref{fig.architecture_paper_subset}, the Inference Engine consists of a Data Structuring Unit and a Feature Computation Unit.

\noindent\textbf{Pre-processing: Octree construction \& memory }
\textbf{pre-configuration}
For each frame of the point cloud, an Octree is constructed in the CPU according to the raw point cloud data and configured into an Octree-Table. The Octree can be built by traversing points in the raw point cloud in a single pass of the data. Additionally, the point cloud data in the Host Memory will be configured (i.e., reorganized) according to the Octree-based sequence. The Octree-Table will record the addresses of the re-arranged points. More details of this process and how it greatly reduces Host Memory access will be given in Section \ref{Pre-processing engine}. After the Octree-Table is built, it is transferred to the Down-sampling Unit in the FPGA.

\noindent\textbf{Preprocessing: downsampling. }In Section \ref{Pre-processing engine}, we will introduce an Octree-Indexed-Sampling (OIS) method which will make use of the Octree-Table \cite{madeira2009gpu} to perform the down-sampling process on memory addresses without having to access the Host Memory, greatly reducing the memory-intensive. As shown in Figure~\ref{fig.architecture_paper_subset}, this down-sampling step is performed in the FPGA in the Down-sampling Unit. The output of the down-sampling is recorded in the Sampled-Points-Table (SPT), which contains the addresses of after-sampled points. Using these addresses in the SPT, the Down-sampling Unit can read the after-sampled points directly from Host Memory and these points provide the input for the inference phase.

\noindent\textbf{Inference Engine. }The Inference Engine consists of two main modules, the Data Structuring Unit (DSU) and the Feature Computation Unit (FCU), both of which are implemented in the FPGA. In our prototype implementation, the FCU is a commercially available Deep Learning Accelerator (DLA), like the Intel NPU \cite{song20197}. The DLA will perform inference on the point-subset as the “input feature map”, which is the output from the DSU. 

The Data Structuring Unit (DSU) performs a preparation step to produce an input feature map that can be used by existing DLAs. The implementation of the DSU is based on a Voxel-Expanded Gathering method and fundamentally reduces the workload of the data structuring step as compared to current methods. The details of the DSU and the Voxel-Expanded Gathering method will be presented in Section \ref{Inference engine}.
  \begin{figure*}[t]
\centering
		\includegraphics[width=1\linewidth]{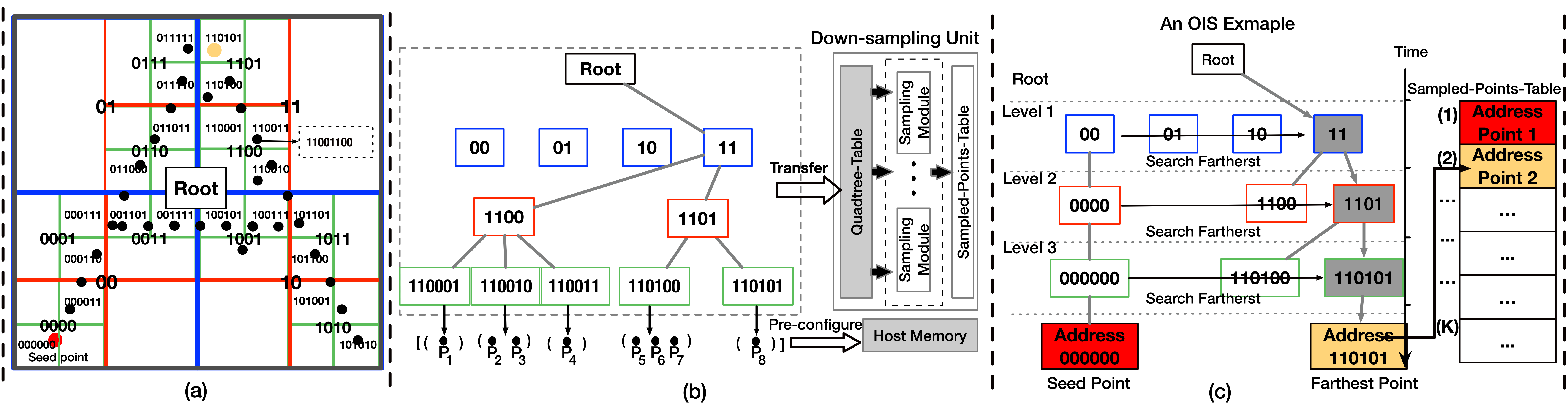}
        \caption{Octree-Indexed-Sampling method overview: (a) A point cloud character “A” (black and color points). Note, for simplicity, that it is a 2D Quadtree illustration of our Octree-Index-Sampling (OIS) method. An Octree contains two horizontal levels of Quadtrees, having an extra Z dimension. (b) Corresponding Quadtree representation. For simplicity, only node “11” is fully expanded. As shown, the content of a Quad-tree is stored in a Quadtree-Table in the Down-sampling Unit; and the raw points corresponding to the Quadtree are pre-configurated in the Host Memory. (c) An example of OIS steps to create the Sampled-Point-Table, which contains the corresponding Host Memory addresses of the K picked points, where K is a pre-defined number, e.g., 4096.}
\label{fig.SWOA_d_pre-eng_blackoutline}
\end{figure*} \normalsize

\section{Pre-processing Engine}

\label{Pre-processing engine}
As we discussed in Section \ref{Motivation}, the pre-processing of a point cloud frame using standard methods is a memory-intensive task. An extremely large amount of input points and intermediate data need to be accessed repetitively from memory; and as a result, the computational workload for pre-processing is also extremely heavy. In this section, we introduce an Octree-Indexed-Sampling (OIS) method to greatly reduce the memory-intensive bottleneck in the down-sampling process. With the OIS method, the HgPCN builds the Octree Table in the Octree-build Unit in the CPU and transfer it to the FPGA via MMIO. Based on the Octree Table, the down-sampling process is performed in the Down-sampling Unit in the FPGA, which directly access the desired after-sampled points from memory. Also in this section, we present how the Down-sampling Unit accelerates the large computational workload through hardware parallelism.

\noindent\textbf{\underline{Octree-Index-Sampling (OIS)}}
Our OIS method is based on a key observation: down-sampling can be based on the relative position of a point in a 3D space. For example, most of the down-sampling methods normalize the point cloud data (to produce the relative positions) before the actual down-sampling process. In the OIS method, we will describe how we can obtain the relative position information beforehand by using a spatial indexing method. Then we can directly use the index to perform the down-sampling without having to access the memory to read out the absolute XYZ coordinates.

\subsection{Octree-build Unit in the CPU}
\noindent\textbf{Octree Construction.} 
The Octree structure is a way to regularize the point cloud using voxels. It decomposes the point cloud distributed in a 3D space into a 1D-array, where the leaf level of the generated Octree is the resulting 1D-array. Figure~\ref{fig.SWOA_d_pre-eng_blackoutline}(a) illustrates the steps of creating a Quadtree (note that a Quadtree, a simplified 2D-version of an Octree, is used for illustration). In the beginning, we put the point cloud (e.g., point cloud character “A”) into a root-level voxel (the outer black bordered “box” in Figure~\ref{fig.SWOA_d_pre-eng_blackoutline}(a)). Then we continuously divide each non-empty voxel into sub-voxels (4 “blue” sub-voxels for the example Quadtree in Figure~\ref{fig.SWOA_d_pre-eng_blackoutline}(a); 8 sub-voxels in an actual Octree) until it reaches a pre-defined depth (root plus three levels in this example). For each division, the level of the Octree is increased by one and the m-codes \cite{morton1966computer} of the subdivided nodes add two more bits at the end (three more bits in the Octree). In these newly added bits, the first bit represents the X-axis, and the second bit represents the Y-axis (for an Octree, the third bit represents the Z-axis).

As shown in Figure~\ref{fig.SWOA_d_pre-eng_blackoutline}(a), these new-added bits are based on the Space-Filling Curve (SFC) traversal order \cite{sakoglu2020adaptive}. For the sub-voxels (child-nodes) generated by each subdivision, the new-added code of each sub-voxel, when compared with its parent voxel, is based on its relative position inside the parent voxel: i.e., the bottom-left quadrant is 00; the top-left is 01; the bottom-right is 10, and the top-right is 11. 

Shown in Figure~\ref{fig.SWOA_d_pre-eng_blackoutline}(b) is a partial representation of the resulting Quadtree. The root and three levels are color-coded, corresponding to Figure~\ref{fig.SWOA_d_pre-eng_blackoutline}(a). Every non-empty voxel from Figure~\ref{fig.SWOA_d_pre-eng_blackoutline}(a) will be represented by a node in the Quadtree in Figure~\ref{fig.SWOA_d_pre-eng_blackoutline}(b). In the final generated Quadtree, each non-leaf node includes up to four child-nodes (sub-voxels). Each leaf node includes the actual points within that node. For example, the leaf node 110011 shown in Figure~\ref{fig.SWOA_d_pre-eng_blackoutline}(a) includes one point (11001100) and the leaf node 001101 contains four points.

As described in \cite{schon2013octree}, Octree can be used as an indexing method in a \textit{spatial} database to optimize \textbf{spatial queries}. In HgPCN, the spatial information of each point is stored in Host Memory and can be obtained efficiently through an Octree-Table lookup. To do so, the point cloud data has to be reorganized as described next. 

\noindent\textbf{Octree-based Organization in Host Memory.} 
The Octree construction (as described in the previous section) is a process of mapping the voxels (nodes) from a higher dimension (2D in Quadtree; 3D in Octree) to 1-D linear ordering. This resulting 1D linear ordering can be naturally mapped to consecutive memory addresses, which is also a 1D array. In Figure~\ref{fig.SWOA_d_pre-eng_blackoutline}(b), connecting the leaf nodes together from the left-most leaf node (110001) to the right-most leaf node (110101) is the 1D linear ordering based on SFC traversal. Based on this 1D order, we can construct a 1D array comprised of the points inside the leaf nodes. In cases when a leaf node contains multiple points (such as Node 110010 or 110100), the intra-node point arrangement also follows the SFC traversal. Note again that Figure~\ref{fig.SWOA_d_pre-eng_blackoutline}(b) is only a partial representation of the example Quadtree from Figure~\ref{fig.SWOA_d_pre-eng_blackoutline}(a), thus the 1D array  $[P_1,P_2,…,P_8]$ shown at the bottom of Figure~\ref{fig.SWOA_d_pre-eng_blackoutline}(b) represents only a part of the complete 1D array for this example. The key point is that this 1D-array represents the same points in the raw point cloud, but are in a reorganized sequence.

Next, the resulting 1D array is used to pre-configure the point cloud data in Host Memory by creating a reorganized copy in the memory. Using the simplified example in Figure~\ref{fig.SWOA_d_pre-eng_blackoutline}(b), the points $[P_1,P_2,…,P_8]$ are initially distributed “irregularly” throughout the million-point dataset in Host Memory. After the Octree-based reorganization process, those points $[P_1,P_2,…,P_8]$ are stored in consecutive addresses in the Host Memory. 

In summary, the Octree construction and the point cloud data reorganization are performed at the same time by a single pass of raw point cloud data. As will be described in the next subsection, through an Octree-Table lookup, we can obtain the memory address to directly access the spatial information (or characteristics) of the desired point. In contrast, the commonly used FPS method has to read out every point in the raw point cloud in Host Memory and searches for the desired next point by a ranking operation.

\begin{figure}[t]
\centerline{\includegraphics[width= 1\linewidth]{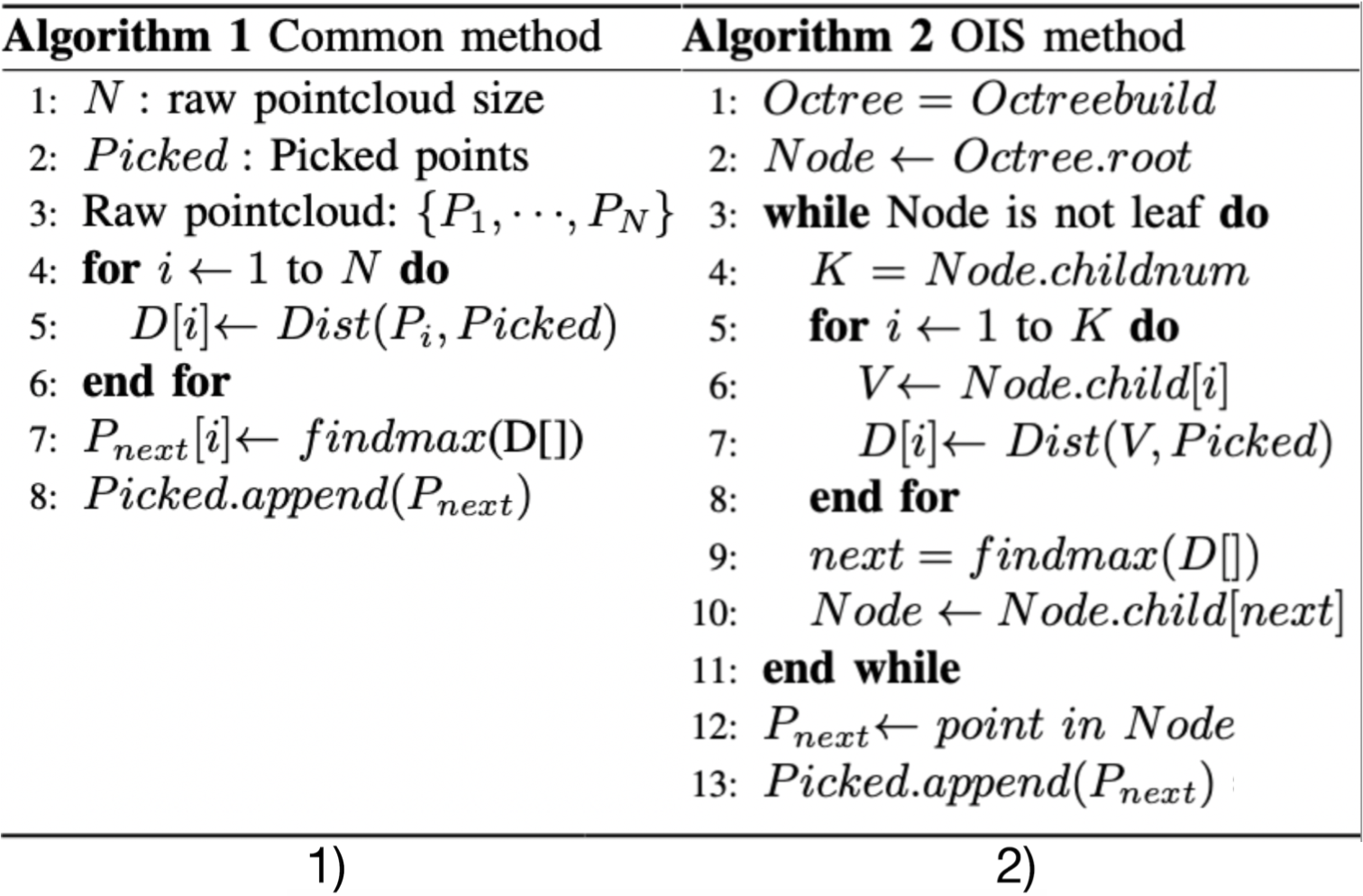}}
\caption{Pseudocodes of 1) common method for FPS. 2) OIS method for FPS.}
\label{fig.pseudocode_final}
\end{figure}

\subsection{Down-sampling Unit in the FPGA}
\label{Down-sampling Unit}

\noindent\textbf{Configuration and use of Octree-Table.} 
As shown in Figure~\ref{fig.SWOA_d_pre-eng_blackoutline}(b), the generated Octree will be configured into an equivalent  Octree-Table, to be transferred to and used by the Down-sampling Unit in the FPGA. In the Octree, the leaf nodes contain the address (or address range) of the contained point(s). In the Down-sampling Unit, the Sampling Modules perform (in parallel) the down-sampling task by performing Octree-Table lookup operations. When a Sampling Module reaches a leaf node during an Octree-Table lookup, the Down-sampling Unit can determine the address containing the desired point and access the Host Memory $directly$ to obtain the spatial information $(x_k,y_k,z_k)$ and feature information $(f_k)$ of that point, as described in the following example.

\noindent\textbf{Octree-Indexed-Sampling Algorithm and an Example.}
Figure~\ref{fig.SWOA_d_pre-eng_blackoutline}(c) shows an example of the steps in using the OIS method to achieve the same function as the commonly-used FPS method, but without incurring the cost of repetitive memory accesses. As illustrated in Algorithm 1 in Figure~\ref{fig.pseudocode_final}, for picking the farthest point as the next picked point, the common FPS method searches every point in unpicked point set $(C - S)$ and compares distances. The OIS method substitutes the operations of Algorithm 1, which work on sparse points, with Octree-Table lookup operations. In the OIS method (Algorithm 2 in Figure~\ref{fig.pseudocode_final}), the step that picks the farthest point is first approximated by picking the farthest voxel. Then within that voxel, pick the farthest of the points inside this voxel according to the SFC (Space-Filling Curve) traversal order. In Algorithm 1, the for-loop for finding the next point (Line 4 to Line 6) iterates N times (N is the number of points in an up-to-million point raw point cloud frame). In contrast, the while-loop in the Algorithm 2 for finding the next point (Line 3 to Line 11) iterates at most a number equal to the depth of the Octree, which is quite limited.

For example, assume that we pick the red point (left bottom-most point in Figure~\ref{fig.SWOA_d_pre-eng_blackoutline}(a)) as the seed point. As the first picked point, the seed point is written into the first entry of the Sampled-Point-Table, as shown in Figure~\ref{fig.SWOA_d_pre-eng_blackoutline}(c). Since this seed point belongs to voxel 000000, we want to find the farthest voxel from 000000. The process is explained as follows.

In the Octree (Figure~\ref{fig.SWOA_d_pre-eng_blackoutline}(c)), the seed voxel 000000 belongs to voxel 00 in the first level; 0000 in the second level; and 000000 in the third (leaf) level. We begin to search for the desired farthest voxel of the seed voxel at the first level. At the first level, voxel 000000 belongs to voxel 00 and its farthest first-level voxel is 11. (Note that the distance between two voxels can be determined by the Hamming distance \cite{norouzi2012hamming} between m-code \cite{hunter2020using}). Voxel 11 has two child-voxels in the second level, 1101 and 1100, of which 1101 is the farther voxel from the 0000. Continuing, voxel 1101 has two child-voxels, 110100 and 110101, of which 110101 is the farther voxel from 000000. Now we reached the leaf and find voxel 110101 as our desired farthest voxel from the seed voxel 000000. For the points inside voxel 110101, we pick the farthest point according to SFC traversal sequence and insert it into the next entry of the Sampled-Point-Table, as shown in Figure~\ref{fig.SWOA_d_pre-eng_blackoutline}(c). 

Continuing, in the standard FPS method \cite{eldar1997farthest}, when the picked points set $S$ contains more than one point (two points in this example thus far), the seed point for the next iteration will be the Euclidean norm of $S$ (i.e., a virtual summary point to represent $S$), denoted by $||S||_2$. The OIS method follows the same approach and repeats the steps of picking the subsequent farthest points from $||S||_2$ until the size of the point set N(S) reaches the pre-defined number $K$.

\begin{figure}[t]
    \centerline{\includegraphics[width= 1\linewidth]{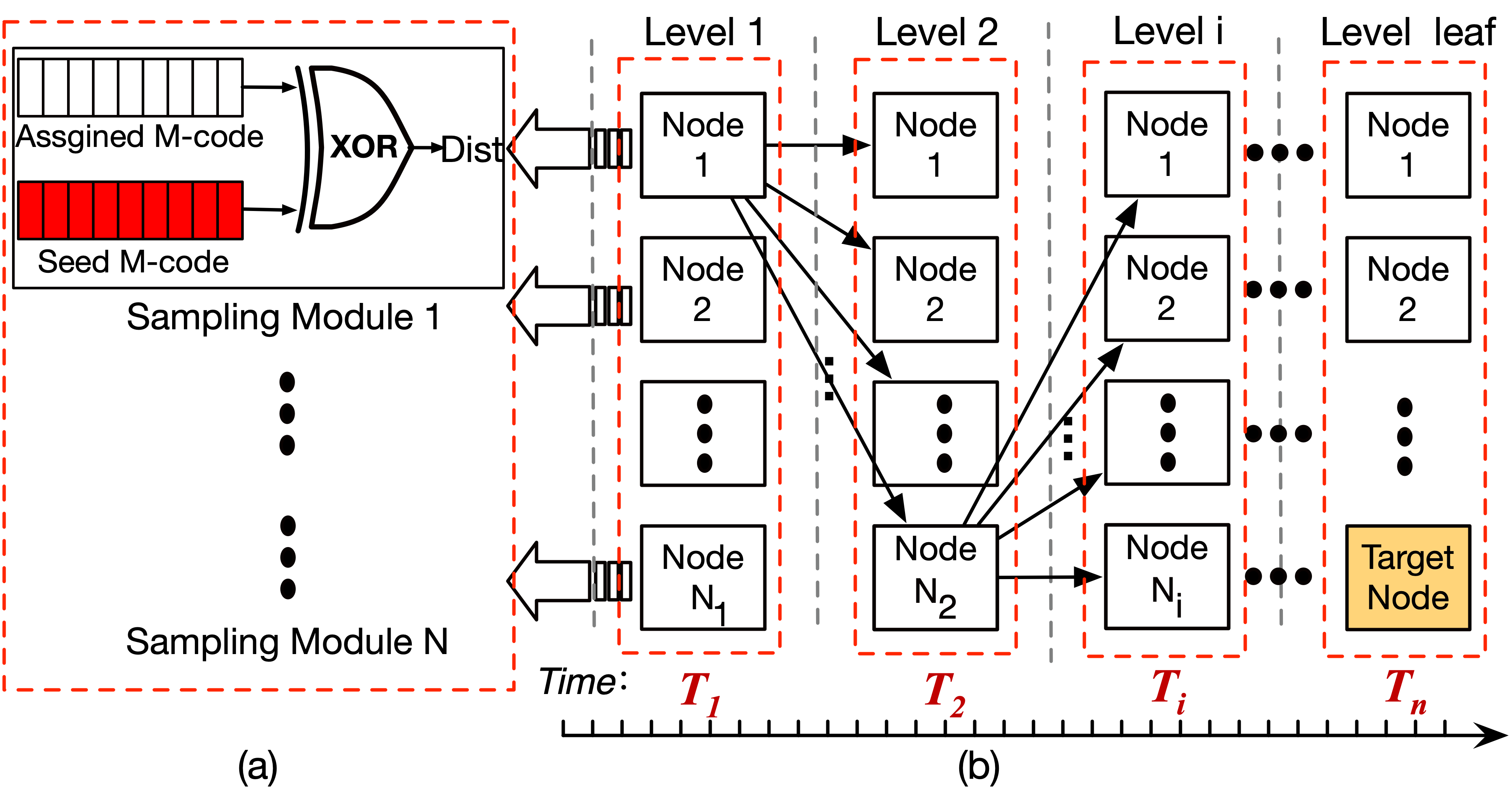}}
    \caption{Hardware design of voxel-level parallelism.}
    \label{fig.Voxel level parallelism}
\end{figure}

\noindent\textbf{\underline{Hardware Acceleration}}
Figure~\ref{fig.Voxel level parallelism}(a) shows the details the Sampling Modules of the Down-sampling Unit in HgPCN. Each Sampling Module has two inputs: m-codes of the assigned voxel and m-codes of the seed voxel. 
A Sampling Module calculates the Hamming distance between the two m-codes using an efficient XOR operation \cite{norouzi2012hamming}.
The outputs from the Sampling Modules are simultaneously inputted into a bitonic sorter (not shown) to select the node with the largest Hamming distance as the next searched node.

To further accelerate the OIS-based down-sampling phase, the Down-sampling Unit deploys multiple Sampling Modules based on \emph{voxel-level parallelism}. Figure~\ref{fig.Voxel level parallelism}(b) shows the steps of OIS-based FPS using voxel-level parallelism. At any given time, eight Sampling Modules are each assigned one of these child-nodes to find the farthest node in parallel.

\noindent\textbf{Summary:}
In summary, to efficiently perform point cloud down-sampling, HgPCN employs an Octree-Indexed-Sampling (OIS) method by utilizing the Octree-based \textit{spatial queries} to directly access the target points in Host Memory. To do so, HgPCN first constructs an Octree-Table and performs a pre-configuration step by reorganizing the point cloud data in the Host Memory. In our prototype HgPCN, these two processes are performed simultaneously in the Octree-build Unit (in the CPU) with a single pass of the raw point cloud data. After that, the Down-sampling Unit of HgPCN (in the FPGA) will perform the down-sampling by directly obtaining the memory addresses of the desired after-sampled points from the Octree-Table. In this manner, HgPCN greatly reduces the \textit{memory-intensive bottleneck of the pre-processing phase} by indexing the 3D relative position of points with the 1D memory address. To further accelerate the OIS-based down-sampling phase, HgPCN exploits voxel-level parallelism by using multiple Sampling Modules in the Down-sampling Unit. 

Finally, the OIS method also provides a significant saving of on-chip memory in the FPGA. This is important because it allows the FPGA to have enough on-chip memory to support the accelerators for the two phases in HgPCN (down-sampling and inference) within one device. In Section \ref{Evaluation}, we will evaluate and analyze of the on-chip memory-saving benefit provided by the OIS method.

    \begin{figure*}[t]
	\centering
	\includegraphics[width=1\linewidth]{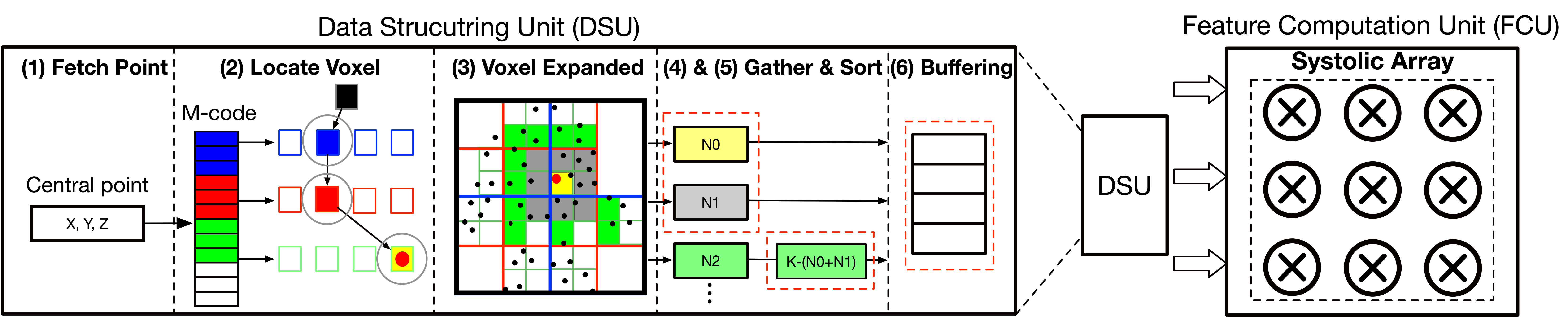}
	\caption{Inference Engine overview.}
	\label{fig.DSU+FCU_CHANNEL}
    \end{figure*} \normalsize  

 \section{Inference engine}

 \label{Inference engine}
As we discussed in Section \ref{Motivation}, for point cloud inference, before the actual feature computation step in the Inference Engine, an expensive data structuring step is required to form an input feature map; without which the inference cannot be supported by current commercially available DLAs.  Thus, as shown in Figure~\ref{fig.DSU+FCU_CHANNEL}, the HgPCN Inference Engine consists of a Data Structuring Unit (DSU) and Feature Computation Unit (FCU). The FCU is a commercially available Deep Learning Accelerator (DLA) which implements a classic systolic array design. The Data Structuring Unit (DSU) is a custom-designed module to optimize the expensive data structuring step and will be described below.

\noindent\textbf{\underline{Data Structuring Unit}}
Most existing PCN accelerators \cite{9251968, lin2021pointacc} accelerate this step mainly from a hardware perspective, e.g., using parallel execution. In this paper, we propose a \textit{Voxel-Expanded Gathering} (VEG) method, in which optimization is achieved through algorithm and hardware co-design. 
The VEG method utilize the \emph{spatial indexing} nature of Octree to algorithmically minimize the required computational workload in the data structuring (DS) step, and then further accelerate through parallel execution.
Furthermore, the VEG method can efficiently support commonly used DS methods, e.g., KNN (K-Nearest-Neighbors) and BQ (Ball Query). In this section, we will use KNN as the example.

\noindent\textbf{Traditional Method for Input Feature Map.} After the down-sampling step, current PCNs typically pick a fixed number of central points (e.g., red points in the input point cloud of Figure~\ref{fig.background_pre_final}) from the down-sampled input point cloud. The data structuring step is used to form the subset of points as the “input feature map” by gathering nearby points of these central points. For the data structuring step of each central point, current PCN accelerators search all the points from a (down sampled) input point cloud, and then calculate the distance from every other point to the center point. Next, they sort and pick out the nearest K neighbor points from input point cloud. For example, if we assume the down-sampled input point cloud (after the processing phase) has 4096 points, and KNN needs to gather the $K=32$ nearest points from the central point. With traditional methods, they need to compute 4095 distances from the central point to every other point, and select the top $K=32$ points with shortest distances. In this process, most of the 4095 distance calculation (except for the 32 nearest neighbors) can be regarded as wasted.

\noindent\textbf{Voxel-Expanded Gathering (VEG) Method.} In the VEG method, before the actual sorting step, we first narrow the range of nearest point search by \emph{adjacent-indexing} through the use of an Octree. The standard Octree neighbor-search operation \cite{frisken2002simple} is used to search the voxels adjacent to a central point’s voxel. 
More details of the steps of VEG method are discussed below. The HgPCN Data Structure Unit can execute multiple Octree neighbor search operations in parallel to search all adjacent voxels at the same time.

\noindent\textbf{Architectural Support In HgPCN.}  
As shown in Figure~\ref{fig.DSU+FCU_CHANNEL}, the design of the HgPCN Data Structuring Unit consists of six pipeline stages:
\begin{itemize}
\item \textbf{1) Fetch Central Points (FP)}: Fetch a central point and the corresponding m-code \cite{morton1966computer} of this point.

\item \textbf{2) Locate Central Voxel (LV)}: Locate the voxel that contains the central point $V_{seed}$ (yellow voxel shown in Figure~\ref{fig.DSU+FCU_CHANNEL} (2 and 3)).

\item \textbf{3) Voxel Expansion (VE)}: At the first level of voxel expansion ($V1$), neighboring voxels touching the central voxel $V_{seed}$ are included (the gray voxels in Figure~\ref{fig.DSU+FCU_CHANNEL} (3)). Voxel expansion continues outward and include the next level of touching voxels (the green voxels are included for the second expansion $V_2$). Voxel expansion continues until the total number of points included in the expanded voxels is at least $K$.
 
\item \textbf{4) Gather Points (GP)}: Let’s assume that there are $N_{0}$ points in the initial voxel $V_{seed}$ and that we need n expansions to collect enough points ($>=K$). Also, assume there are $N_1$ points in gathered voxels by first expansion $V_1$; and $N_n$ points in gathered voxels by the n-th expansion $V_n$ . Gather the points from $V_{seed}$ to $V_{n-1}$ (i.e., total of $N_{0}+N_1...+N_{n-1}$ points).
 
\item \textbf{5) Sort (ST)}: Sort the points in last voxel set $V_{n}$ and pick the $K$ - sum($N_{0}$ to $N_{n-1}$) points with nearest distances from the central point. Together with the $N_{0}+N_1...+N_{n-1}$ points, the $K$ nearest points are found.

\item \textbf{6) Buffering (BF)}: Output the gathered K points, including the coordinates $(p_{k})$ and feature information $(f_{k})$, to the input buffer for the feature computation step. 

\end{itemize}

The above steps are repeated for each central point. 

\noindent\textbf{Example:} 
Assume we need to gather the $32$ nearest neighbors for the central point (red point) in Figure~\ref{fig.DSU+FCU_CHANNEL}.
\begin{itemize}

        \item The voxel (yellow) that contains the red point is the seed voxel $V_{seed}$. Let’s assume there are $N_{0}$ points gathered in $V_{seed}$ and that $N_{0} <32$.

\item The voxel search is expanded outward from $V_{seed}$ and includes first-level neighbor voxels of $V_{1}$ (grey voxels in Figure~\ref{fig.DSU+FCU_CHANNEL}). Assume there are $N_{1}$ points in the set of grey voxels and that $N_{0} + N_{1} < 32$. These $N_{1}$ points are gathered in  $V_{1}$.
      
\item Since there are still $<32$ in the gathered voxels, we continue to include the second-level neighbor voxels (green voxels). Let’s assume there are  $N_{2}$ points in the set of green voxels. These $N_{2}$ points are gathered in $V_{2}$.

\item Let’s assume that $N_{0}+N_{1}+N_{2}>32$, then pick the top nearest $(32-N_{0}-N_{1})$ points in $V_{2}$ and combine them with $N_{0}$ and $N_{1}$ to form the 32 nearest neighbors.
\end{itemize}

 Note that it is unnecessary to sort or compute the distances of points in $N_{0}$ (from $V_{seed}$) and $N_{1}$ (from $V_{1}$), because they are definitely among the 32 nearest neighbors. Existing methods usually compute the distance from the central point to every other point in the 4096 input points, and pick the 32 nearest points. With the VEG method, the computation workload (of sorting and picking the nearest points) is reduced from 4095 to $N_{2}$. 

\section{Evaluation}
\label{Evaluation}

\subsection{Evaluation Setup}

\noindent\textbf{Implementation Method}
A prototype of HgPCN has been implemented on an Intel PAC, a shared memory CPU-FPGA (Intel Arria 10 GX) platform. The accelerators are implemented on the FPGA side with SystemVerilog and VHDL, and the software implemented on the CPU side with C++.

\noindent\textbf{Benchmark datasets}
As shown in Table \ref{table.Benchmarks}, four common point cloud datasets (for four applications), with different raw-dataset frame sizes, were selected as our benchmarks.

\begin{table}[h]
\caption{Evaluation Benchmarks}
\centering
\resizebox{0.48\textwidth}{!}{
\begin{tabular}{c|c|c|c} % 控制表格的格式
    \hline
%\textbf{Application}&\textbf{Dataset}&\textbf{Scene}&\textbf{{Raw Size}&\textbf{Model} \\
\textbf{Application}&\textbf{Dataset}&\textbf{input Size}&\textbf{PCN Model} \\
    \hline
    Object Classification& ModelNet40 & $1024$& Pointnet++(c)\cite{qi2017pointnet++} \\
    \hline
    Part Segmentation& ShapeNet  & $2048$& Pointnet++(ps)\cite{qi2017pointnet++} \\
    \hline
    Indoor Segmentation& S3DIS & $4096$& Pointnet++(s)\cite{qi2017pointnet++} \\
    \hline
    Outdoor Segmentation & KITTI & $16384$ & Pointnet++(s)\cite{qi2018frustum} \\
    \hline
\end{tabular}}
\label{table.Benchmarks}
\end{table}

\noindent\textbf{Baseline devices and accelerators for comparison}
General-purpose devices used in our comparison include the Intel® Xeon® W-2255, Nvidia Jetson GPU, and 4060ti GPU. These general-purpose devices are used for end-to-end comparison, including the pre-processing phase and the inference phase.

The second type of baseline hardware for comparison is against existing PCN accelerators, Mesorasi \cite{9251968} and PointACC \cite{lin2021pointacc}, with $16$ $\times$ $16$ systolic arrays for the feature computation step. Because these PCN accelerators do not include the pre-processing phase (i.e., not end-to-end), we will perform the comparison only on the PCN inference phase.

\subsection{Analysis of the OIS method on CPU}
\begin{figure}[t]
    \centering
    \includegraphics[width= 1\linewidth]{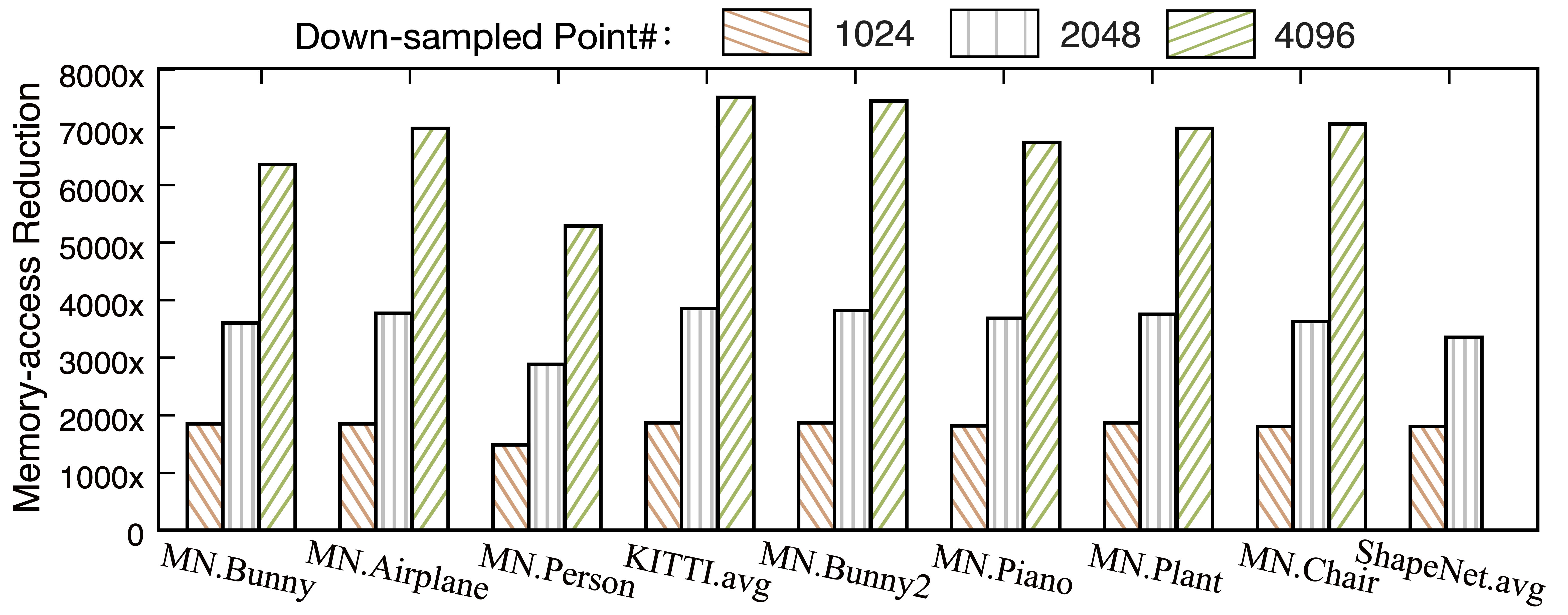}
    \caption{Memory-access saving from OIS-method.}% over different size point cloud frames}
    \label{fig:SWOA_memorysaving}
\end{figure}
As stated, for the FPS down-sampling algorithm, the OIS method converts original operations (Algorithm 1 in Figure~\ref{fig.pseudocode_final}) on sparse points into efficient Octree lookup operations (Algorithm 2), optimizing the memory-intensive problem of the pre-processing phase. To demonstrate the inherent advantage of the OIS method against the common FPS method, we first evaluated these algorithms running on the same CPU (for OIS, both Octree-build Unit and Down-sampling Unit are on CPU).

\begin{figure}[t]
    \centering
    \includegraphics[width= 1\linewidth]{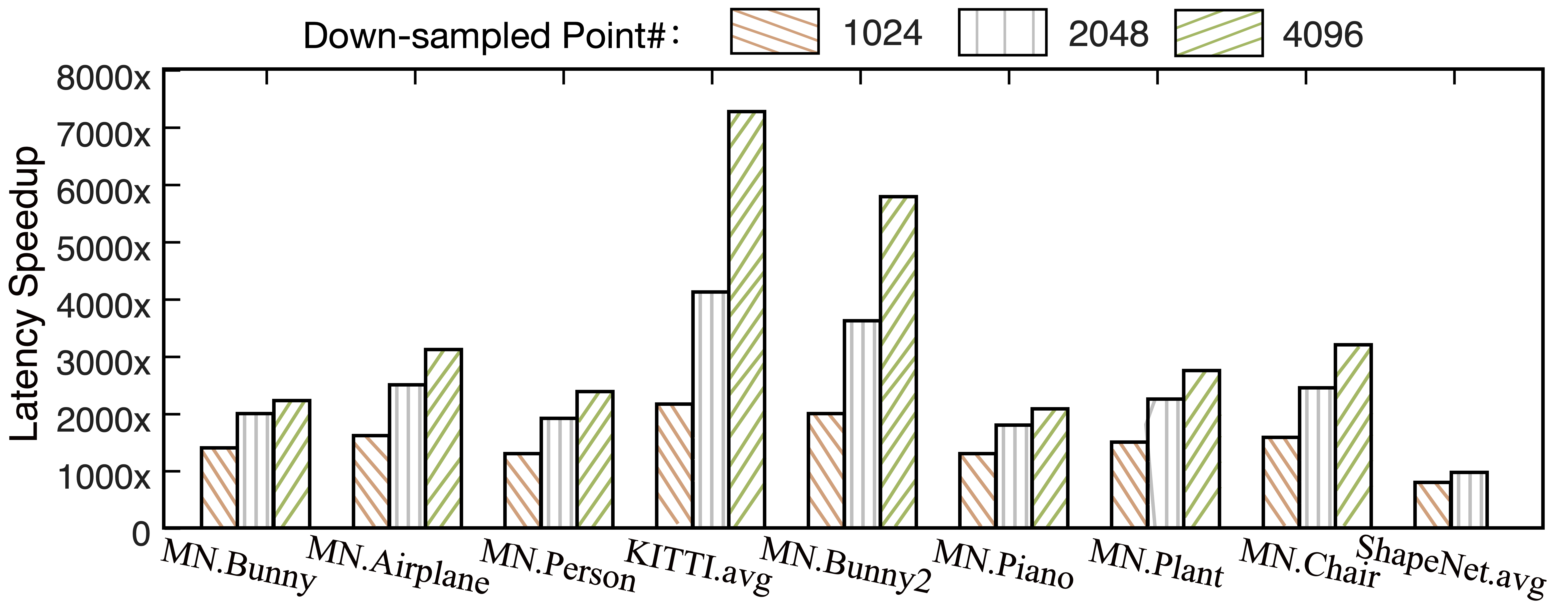} %[scale=0.55]
    \caption{latency speedup from OIS-method on CPU.}% over different size point cloud frames}
    \label{fig:spd_SWOA_CPU_10.16}
\end{figure}

\noindent\textbf{Benefits of the OIS method.}
Both Algorithm 1 (common FPS-method) and Algorithm 2 (OIS method) were run on an Intel® Xeon® W-2255 CPU, using point cloud frames with different sizes. The results are shown in Figure~\ref{fig:SWOA_memorysaving} and \ref{fig:spd_SWOA_CPU_10.16}. In both figures, the x-axis shows the different point cloud frames. Labels beginning with “MN.” are frames from the ModelNet40 dataset and the label “kitti.avg” represent a frame from the KITTI database of the average size. Note that for Shapenet, the raw data size is smaller than 4096 points, so it doesn't include a column for down-sampled to 4096 points. For the y-axis, 6000$\times$ means the OIS method requires 6000 times less memory accesses than FPS method.

As shown in Figure~\ref{fig:SWOA_memorysaving}, memory-access saving for these benchmarks ranges from 1700$\times$ to 7900$\times$ for point cloud frames of different sizes. This result is consistent with the theoretical analysis of memory-access saving between Algorithm 1 and Algorithm 2 in Figure~\ref{fig.pseudocode_final}. 

Figure~\ref{fig:spd_SWOA_CPU_10.16} shows the overall measured latency improvement of the OIS-based method, as a result of the memory-access savings. The OIS-based sampling outperforms the common FPS method with 800$\times$ to 7500$\times$ speedup.

\begin{figure}[htp]
\centerline{\includegraphics[width= 1\linewidth]{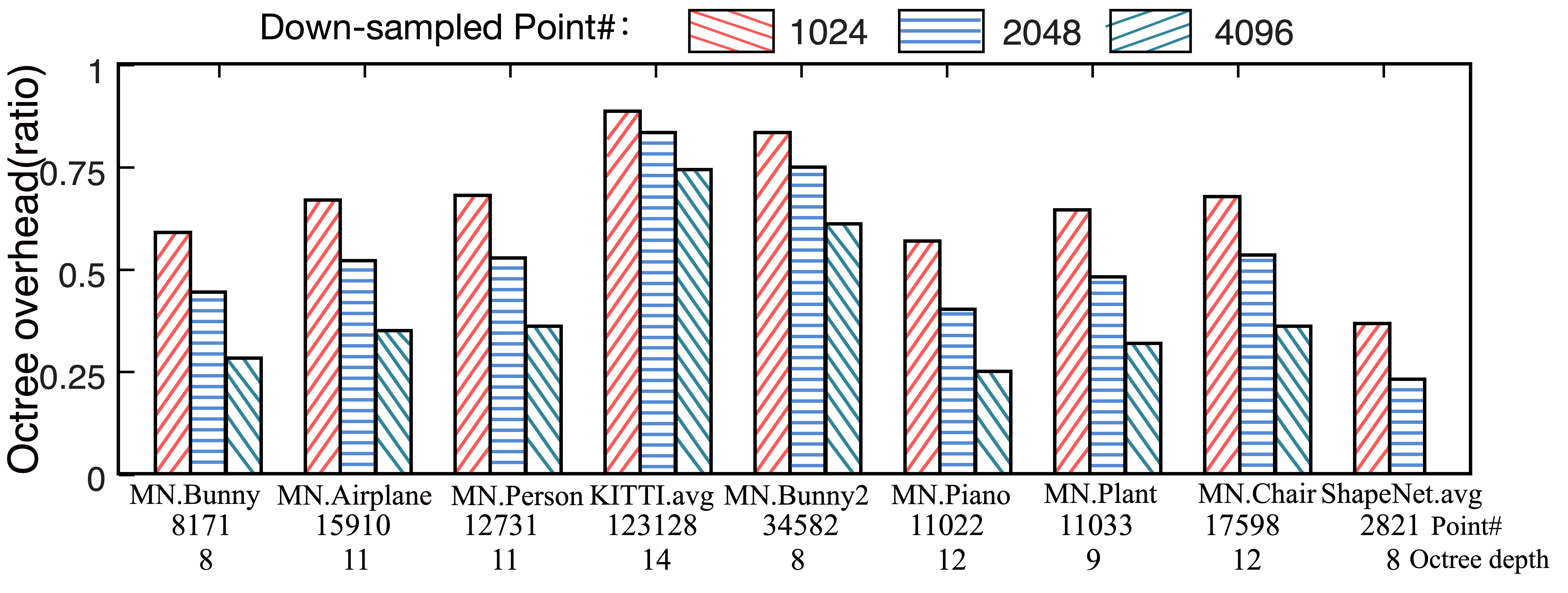}}
\caption{Octree-build overhead of OIS-based sampling.}
\label{fig:octree_overhead}
\end{figure}

\noindent\textbf{Overhead from Octree Build.}
As discussed in Section \ref{Pre-processing engine} (Pre-processing Engine section) and shown in Algorithm 2, the Octree-indexed Sampling method requires an initial overhead of building the Octree. Figure~\ref{fig:octree_overhead} details the overhead for different point cloud datasets and frame sizes. The Octree-construction overhead can range from 0.25 to 0.8 of the total latency of the OIS method when implemented only on CPU. In OIS-based sampling (on CPU), most of the latency and memory accesses are from the Octree-construction, because the OIS method must go through the entire raw point cloud data once to build the Octree. 
Note also, that the latency of OIS is impacted by two factors: the number of points in the point cloud and the depth of the Octree. The Octree-construction latency is determined by the number of points in a point cloud, and the speed of Octree-search is linearly related to the Octree depth. The Octree depth is influenced by the non-uniformity of a point cloud \cite{laine2010efficient}. In Figure \ref{fig:octree_overhead}, even though MN.piano and MN.plant contain almost the same number of points, the spatial distribution of MN.piano is more non-uniform than that of MN.plant, resulting in a deeper Octree for MN.piano.
Unlike traditional methods, which require repetitive searching among the raw point cloud data, OIS can greatly reduce the amount of memory access (as demonstrated in Figure~\ref{fig:SWOA_memorysaving}). Furthermore in HgPCN, the VEG method (for DS step) can reuse the built Octree to amortize the overhead.

\subsection{Pre-processing Engine Performance}

\begin{figure}[htp]
    \centering
    %\centerline{\includegraphics{5sampling.png}}
    \includegraphics[width= 1\linewidth]{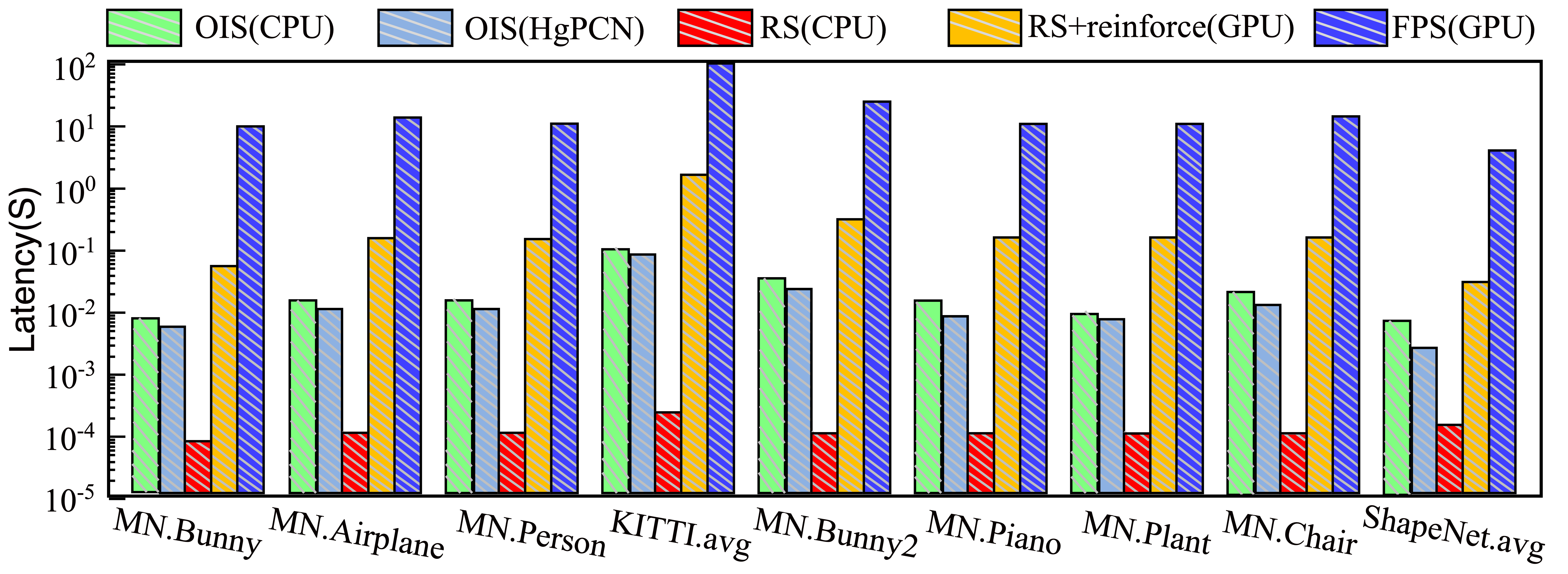}
    \caption{Latency comparison of Pre-processing Engine against the baselines}
    \label{fig:5sampling_method}
    \end{figure}

In Figure~\ref{fig:5sampling_method}, we first compare the OIS-on-CPU (software implementation on CPU) vs OIS-on-HgPCN to illustrate the speedup resulting from the hardware Down-sampling Unit. As seen from the first two columns in Figure~\ref{fig:5sampling_method}, compared with OIS-on-CPU, the OIS-on-HgPCN can provide a from 1.2$\times$ to 4.1$\times$ speedup. This speedup stems from the reason that the hardware Down-sampling Unit (described in Section \ref{Down-sampling Unit}) can achieve a 5.95$\times$ to 6.24$\times$ speedup compared to the CPU-implemented Down-sampling Unit.

In the remainder of Figure~\ref{fig:5sampling_method}, we compared the runtime latency of HgPCN’s Pre-processing Engine with three existing sampling methods on the baseline devices where they can get the best performance (CPU or GPU). (To our best of knowledge, there is no existing DSA for point cloud pre-processing). Among these sampling methods, FPS is the most commonly used because it results in the least information loss. However, it has the highest runtime latency. Random sampling (RS) has the lowest runtime latency, but is not generally used because it has the highest information loss. In some existing methods \cite{hu2020randla}, a reinforcement process uses an encoder architecture to enhance RS (as RS+reinforce). But this method is not universal. It can only be applied to PCNs with encoder-decoder architectures.

In Figure~\ref{fig:5sampling_method}, the total latency of the Octree-indexed-sampling (OIS) method on HgPCN (light-blue columns) is shown. This latency includes the CPU-end (Octree-build Unit) and the FPGA-accelerator-end (Down-sampling Unit). The results show that the OIS method on HgPCN possesses the best advantages of the tested sampling methods. Although the performance of OIS on HgPCN is a little slower than random sampling, it can achieve the same accuracy as the FPS method. Unlike RS+reinforce, it is universal to all types of point cloud networks. Furthermore, compared to the FPS method, HgPCN offers a more consistent latency for different sizes of point cloud frames, providing better tail latency for edge computing.

\begin{figure}[htp]
\centerline{\includegraphics[width= 1\linewidth]{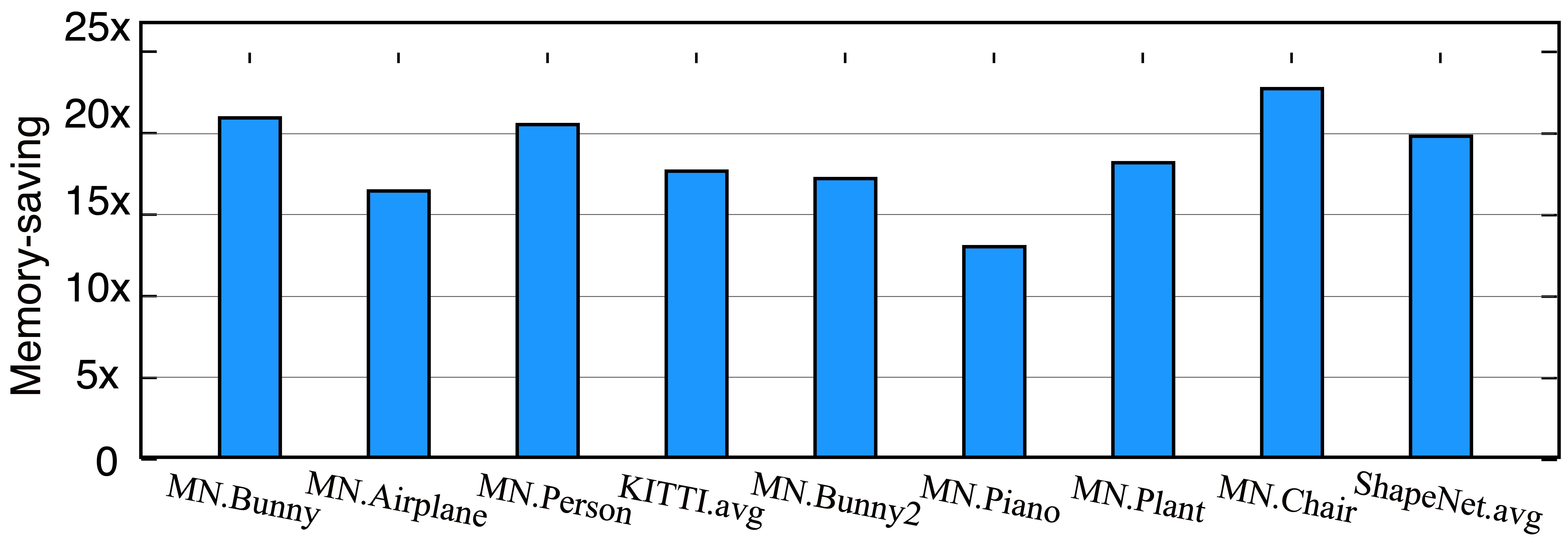}}
\caption{On-chip memory-saving benefit from the OIS method.}
\label{fig:onchip-memory-saving}
\end{figure}

 \noindent\textbf{On-chip memory-saving with the OIS method.} Reducing on-chip memory usage in an FPGA is important: 1) to free up resources to implement other parts of the application and 2) to have the opportunity to keep an entire implementation on a single FPGA. Figure~\ref{fig:onchip-memory-saving} shows the memory-consumption comparison between the FPS method and the OIS method: OIS can provide from 12$\times$ to 22$\times$ memory-saving.

In our prototype implementation of HgPCN, the device used is the Intel Arria 10 GX 1150 FPGA, which has 65Mb total on-chip RAM. From our evaluation, if the raw point cloud frame contains more than ${\sim}5\times10^5$ points and is stored in the FPGA memory (as in the case of FPS), the input points and generated intermediate data during the pre-processing phase will exceed the total capacity (65Mb) of on-chip memory. In this case, there will be no space for the Inference Engine. With the OIS method, only the Octree-Table needs to be stored in the on-chip memory, along with minimum amount of intermediate data. Even if raw point cloud frames contain ${\sim}1\times10^6$ points, the OIS method only consumes approximate 10 Mb of memory in the pre-processing phase.

 \subsection{Evaluation of Inference Engine}

\begin{figure}[htp]
\centerline{\includegraphics[width= 0.99\linewidth]{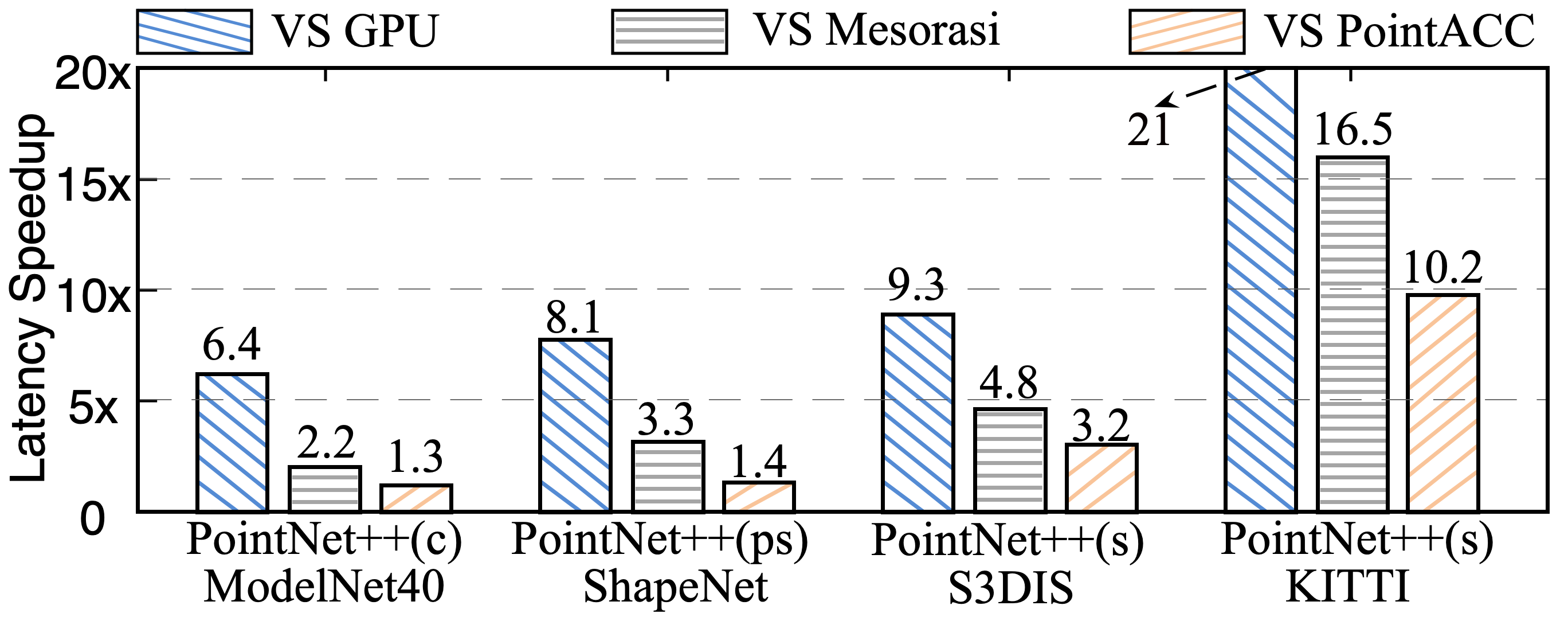}}
\caption{Speedup of HgPCN over baseline hardware. }
\label{fig:Inference_spd}
\end{figure}

In this section, we will present the evaluation results of the HgPCN Inference Engine against a GPU accelerator and two PCN accelerators: PointACC and Mesorasi. Recall that these PCN accelerators do not include the pre-processing phase (i.e., not end-to-end); thus we can only perform the comparison on the PCN inference phase. Also, because the Mesorasi PCN accelerator uses a random picking method to generate central points for the data structuring step, we will do the same for the GPU, PointACC, and HgPCN to ensure a fair comparison. The evaluation results are shown in Figure~\ref{fig:Inference_spd}, in which the x-axis shows the four PCN tasks from Table \ref{table.Benchmarks} and the y-axis shows the speedup of HgPCN against the baselines. 

As shown in Figure~\ref{fig:Inference_spd}, when compared to Nvidia Jetson NX GPU (blue columns), HgPCN achieves from 6.4$\times$ to 21$\times$ speedup. When compared to Mesorasi (grey columns), based on the same systolic array architecture for the feature computation step, HgPCN achieves 2.2$\times$ to 16.5$\times$ speedup. We noted that Mesorasi uses a GPU to perform the data structuring step, which still occupies a major part of the latency. Even though Mesorasi tries to overlap the data structuring and feature computation, the inference speed is still largely limited by the latency of the data structuring step.

\begin{figure}[htbp]
\centering
\begin{minipage}[t]{0.48\linewidth}
\centering
\includegraphics{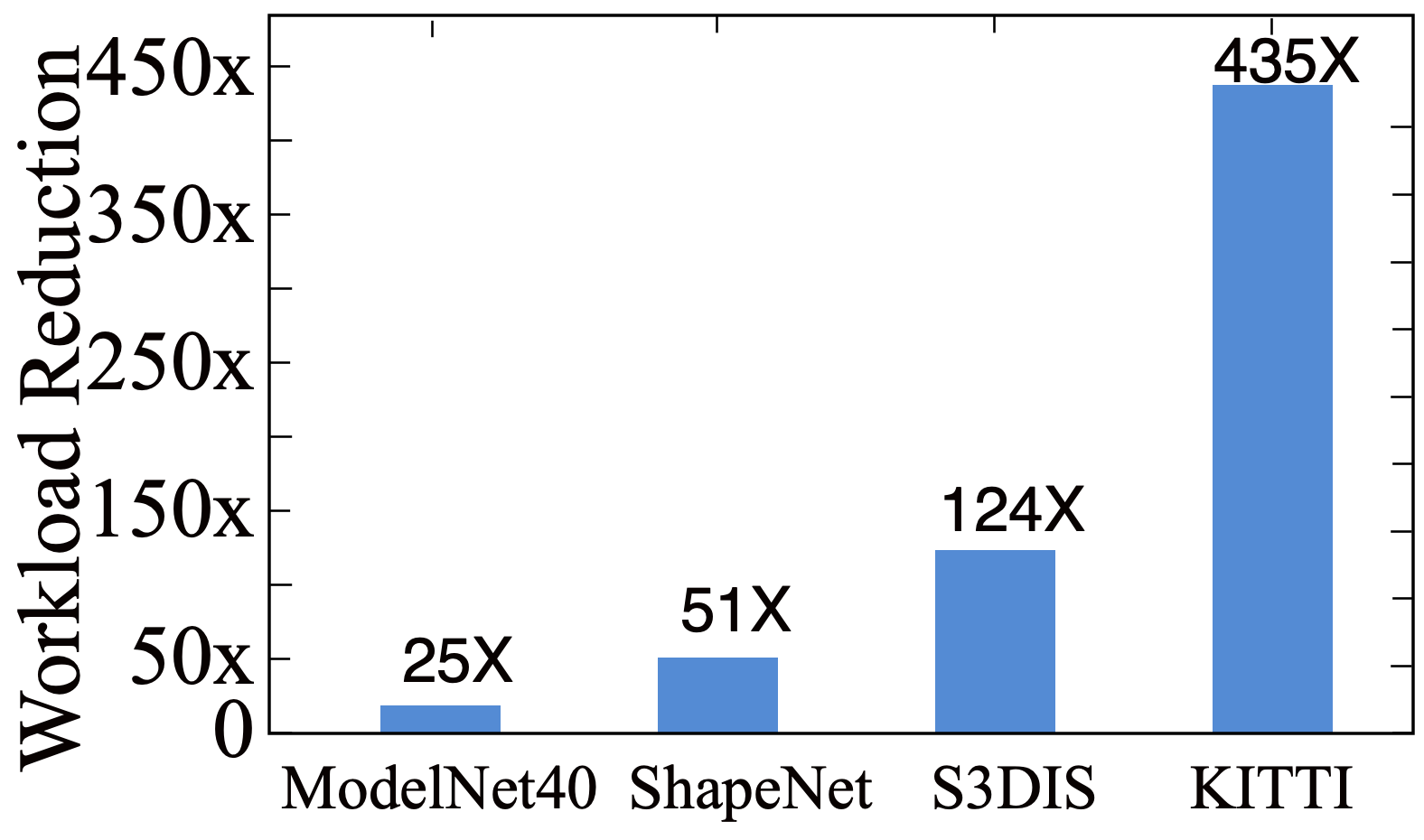}
%\includegraphics[width=2cm]{asplos23-templates-summer-2/input_saving_final.png}
%\caption{Computational Workload reduction }
\caption{VEG benefit.}

\label{fig:input_saving_final}

\end{minipage}
\begin{minipage}[t]{0.48\linewidth}
\centering
\includegraphics{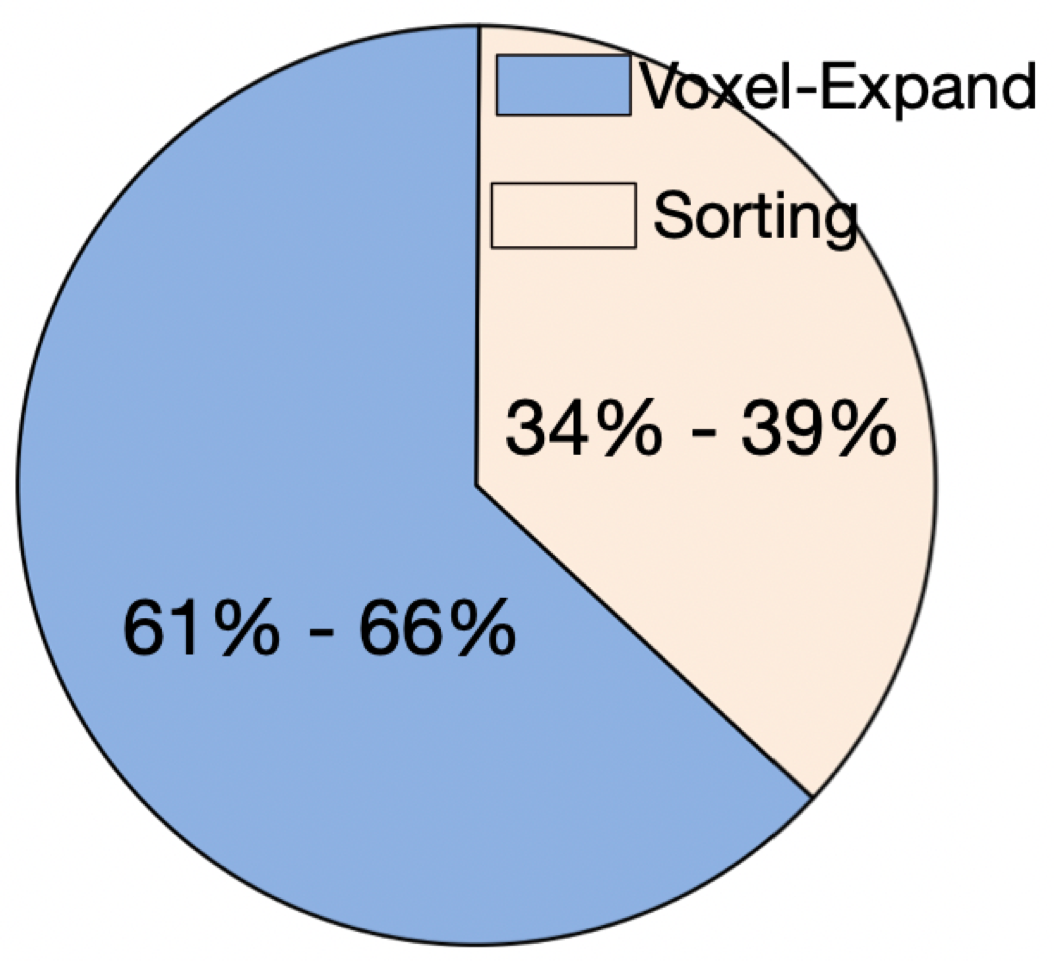}
\caption{VEG breakdown.}

\label{fig:VEG_breakdown}
\end{minipage}
\end{figure}

When compared to PointACC (orange columns), HgPCN achieves 1.3$\times$ to 10.2$\times$ speedup. Like HgPCN, PointACC also developed a customized unit (Mapping Unit) for the data structuring step. Like HgPCN’s DSU, the Mapping Unit of PointACC also uses a bitonic sorter to accelerate the ranking process. However, while based on the same bitonic sorter approach to select the top K nearest points, the HgPCN can outperform PointACC because of the fundamentally reduced workload resulting from the VEG method. Through the VEG method, HgPCN narrows the range for searching the nearest points. In Figure~\ref{fig:input_saving_final}, we compare HgPCN against PointACC to show the benefits of the VEG method. Like other current existing methods, the searched range of PointACC’s bitonic sorter is over the entire input point cloud. In contrast, the required workload of the bitonic sorter in HgPCN’s Data Structuring Unit only sorts the points in $N_{n}$ (from the last expansion $V_{n}$ in Figure~\ref{fig.DSU+FCU_CHANNEL}). Thus, the required workload of the bitonic sorter in HgPCN’s DSU is fundamentally less than that of PointACC’s Mapping Unit. Figure~\ref{fig:input_saving_final} shows the workload reduction due to the benefits of the VEG method. Note that for PCN tasks with larger input sizes, the VEG method can achieve an even greater level of workload reduction. Figure~\ref{fig:VEG_breakdown} shows the latency breakdown of the VEG method.

\subsection{Evaluation of System-level HgPCN}

In previous subsections, we analyzed separately the details of the performance of HgPCN in the pre-processing phase (heterogeneously implemented on the CPU side and the FPGA-accelerator side) and the inference phase (on FPGA side). Here, we combine these two phases to evaluate the E2E system-level performance of HgPCN in an edge computing scenario. For this evaluation, the KITTI dataset is used as our benchmark dataset because each frame of KITTI is associated with a time-label of frame generation, from which we can determine the maximum generation rate of KITTI data frames.

Let’s assume the point cloud frames are generated in real-time frame by frame.
Our definition of meeting real-time requirement is the end-to-end processing of each frame can keep up with the sampling (data generation) rate.
For the KITTI dataset, it was shown that HgPCN can process 16 average frames per second (FPS). For the KITTI dataset, according to the time label of frame-generation, we can determine the maximum generation rate of KITTI data frames is less than 16 frames per second. Thus, we can conclude that HgPCN can keep up with the data generation rate of KITTI to meet the real-time requirement of such an application.

\section{Conclusions and Discussion}
\label{Conclusions}

Latency for end-to-end processing of the point cloud workloads stems from two reasons: memory-intensive down-sampling in the pre-processing phase and the data structuring step for input preparation in the inference phase. In this paper, we presented HgPCN, an end-to-end heterogeneous architecture for real-time embedded point cloud application. In HgPCN, we utilize the space-indexing function of the Octree structure to develop two novel methodologies to address the two identified bottlenecks and accelerate them by using a heterogeneous CPU-FPGA implementation. In the Pre-processing Engine of HgPCN, an OIS method is used to optimize the memory-intensive down-sampling bottleneck of the pre-processing phase. In the Inference Engine, HgPCN extends a commercial DLA with a customized Data Structuring Unit which is based on a VEG method to fundamentally reduce the workload of the data structuring step in the inference phase.

Our results from the previous section showed that for the inference phase, depending on the dataset size, HgPCN achieves speedup from 1.3$\times$ to 10.2$\times$  vs. PointACC, 2.2$\times$  to 16.5$\times$  vs. Mesorasi, and 6.4$\times$  to 21$\times$  vs. Jetson Xavier NX GPU. Along with optimization of the memory-intensive down-sampling bottleneck in pre-processing phase, the overall latency shows that HgPCN can provide the capability of satisfying real-time requirement by keeping up with the raw data generation rate of a benchmark application.

\subsection{Discussion and Future Directions}
\noindent\textbf{Practical Implications} Note that even though both the OIS method in the pre-processing phase and the VEG method in the inference phase are based on the Octree structure, they are independent methods and can be used independently. For example, the HgPCN Pre-processing Engine can be a plug-in to other PCN inference accelerators (not using the VEG method) to perform the end-to-end PCN inference. Similarly, the HgPCN Inference Engine (with VEG method) can be used as an independent PCN inference accelerator. In addition, the OIS and VEG are not only limited to PCN tasks. OIS is applicable to other non-AI point cloud applications (e.g., AR/VR) which require down-sampling of the raw point cloud data. VEG can be used for other point cloud applications \cite{xu2019tigris} which requires neighbor gathering. Also note that the implementation of these two methods for point cloud nets is not limited to a CPU-FPGA platform (as in this paper). The proposed methods are also can be implemented on other accelerators, such GPU and CGRA (Coarse-Grained Reconfigurable Arrays).

Finally, compared to images, the analysis and processing of point cloud data is a relatively unexplored area. The arrangement methods of image pixels (e.g., row-major or column-major) are well-developed, with which a 1D-address can be used as the index to a 2D-coordinate of a pixel. However, to our best of Knowledge, OIS is the first method that can be used to index the 3D-coordinate with 1D-address (by arranging the points). In HgPCN, we demonstrated the benefit of the OIS method and believe it can inspire future works to propose other efficient arranging methods for point cloud.

\noindent\textbf{Future Direction of OIS: Approximate OIS-based FPS.}
Based on OIS, the current prototyped HgPCN accelerates the commonly used down-sampling method, farthest-point sampling (FPS). Considering the approximate nature of DNNs, an approximate OIS-based FPS method can be explored to further enhance the speed while potentially causing only marginal information loss. In the current OIS-based FPS method, the accurate farthest point is identified and added to the sampled point set $S$ by first finding the farthest leaf node and then selecting the farthest point within that node. With the approximate FPS method, instead of finding the accurate farthest point, we can randomly pick a point contained by current accessed node once Octree search is near leaf level. Because the randomly picked point belongs to the same node as the actual farthest point, it is spatially adjacent to the actual farthest point and can serve as an approximate substitute being added to the sampled points set $S$. We will explore the tradeoff between the enhance performance of OIS (by reducing the number of Octree search operations) vs. the lost in accuracy.

\noindent\textbf{Future Direction of VEG: Semi-approximate Data Structuring.}
As we introduced in Section \ref{Background}, unlike the accurate Data Structuring (DS) method used by HgPCN, some PCN accelerators \cite{feng2022crescent, pinkham2020quicknn} opt for the approximate method to accelerate the PCN data structuring step. which requires extra adaption in training. Based on the VEG method, a semi-approximate VEG method can be explored as a future direction, positioned in the middle between accurate and approximate data structuring methods. This semi-approximate VEG method aims to integrate the benefits of both methods.

In the current VEG method, the majority of nearest points are gathered directly from the points included in voxel expansions, except for the final voxel expansion. The points collected in the final voxel expansion need to be sorted to select the rest of the nearest points. This sorting process contributes to most of the computational workload for VEG. With a semi-approximate VEG, the rest of nearest points gathered in the final voxel expansion can be approximated by randomly picking points. Among the results of the semi-approximate VEG, most of the gathered points are accurate, and the points from the final expansion can serve as substitutes for the accurate points because they are spatially adjacent. This way, the semi-approximate VEG can further accelerate the current VEG and potentially without the need for training adaptation.
%%%%%%% -- PAPER CONTENT ENDS -- %%%%%%%%

% \section*{ACKNOWLEDGMENTS}
% This research was supported in part by the National Science Foundation (NSF) Center for Space, High-Performance, and Resilient Computing (SHREC) through the IUCRC Program under Grant No. CNS-1738420.

%%%%%%%%% -- BIB STYLE AND FILE -- %%%%%%%%
\bibliographystyle{IEEEtranS}
\bibliography{refs}
%%%%%%%%%%%%%%%%%%%%%%%%%%%%%%%%%%%%

\end{document}